\pdfoutput=1
\documentclass[a4paper,12pt]{article}

\usepackage[utf8]{inputenc}
\usepackage[T1]{fontenc}
\usepackage{times,graphicx,amssymb, amsmath, comment}
\usepackage{epsfig}%
\usepackage{lmodern}
\usepackage{microtype}
\usepackage{amsmath,amsfonts,amssymb,amsthm}
\usepackage{mathrsfs}
\usepackage[
    colorlinks=true,
    linkcolor=blue,
    urlcolor=blue,
    filecolor=black,
    citecolor=red,
    pdfstartview=FitV,
    pdftitle={},
    pdfauthor={Anjan Sen, Ruchika, Koushik Dutta, Shahin Sheikh-Jabbari, Anirban Roy},
    pdfsubject={},
    pdfkeywords={},
    bookmarksopen=true
]{hyperref}
\usepackage{cleveref}
\usepackage{graphicx}
\usepackage{tensor}
\usepackage{braket}
\usepackage{mleftright} \mleftright
\usepackage{comment}
\usepackage{array}
\usepackage{booktabs}
\usepackage{todonotes}
\usepackage{mathtools} % for mathclap
\usepackage{authblk}
\usepackage{color}
\usepackage{bbold}

\crefname{equation}{}{}

\def\clock{{\count0=\time
           \divide\count0 60
           \ifnum\count0<10 0\fi\the\count0
           \multiply\count0 -60 \advance\count0 \time
           :\ifnum\count0<10 0\fi \the\count0
         }}
\newcommand{\timestamp}{{\small\vbox
{%\hbox{\tt main.tex}
\hbox{\the\day/\the\month/\the\year, \clock}}}}

\usepackage{mathtools}
\usepackage{float}
\usepackage{epsfig}
\usepackage{xcolor, soul}
\sethlcolor{yellow}
\usepackage{hyperref}
\usepackage{booktabs}
\usepackage{hyperref}
\usepackage{cleveref}
\usepackage{color}
\definecolor{darkgreen}{rgb}{0,0.3,0}
\definecolor{darkblue}{rgb}{0,0,0.3}
\definecolor{darkred}{rgb}{0.7,0,0}

\newcommand{\beq}{\begin{equation}}
\newcommand{\eeq}{\end{equation}}
\newcommand{\ber}{\begin{eqnarray}}
\newcommand{\eer}{\end{eqnarray}}
\usepackage{graphicx}
\usepackage{adjustbox}
\usepackage{blindtext}

\def\zm{z_{\text{min}}}

\def\lcdm{$\Lambda$CDM }
\def\rde{{\rho}_{_{\text{DE}}}}
\def\pde{{P}_{_{\text{DE}}}}

\def\beq{\begin{equation}}
\def\eeq{\end{equation}}

\def\ber{\begin{eqnarray}}
\def\eer{\end{eqnarray}}
\begin{document}
%\preprint{ IPM/P-2018/064}
%%%%%%%%%%%%%%%%%%%%%%%%%%%%%%%%%%%%%%%%%%%%%%%%%%%%%%%%%%%%%%%%%%%%%%
%\newcommand{\mytitle}{Cosmological Data Do Not Necessitate Positive Cosmological Constant}
%\newcommand{\mytitle}{Doing Away with the Positive Cosmological Constant}
%\title[Doing Away with the Positive Cosmological Constant]
\title{\textbf{Beyond $\Lambda$CDM with Low and High Redshift Data: Implications for Dark Energy}}
%{Negative Cosmological Constant is Consistent with Cosmological Data}

\vskip 0.5cm
\author[Dutta etal.]{ {Koushik Dutta$^{1,2}$\thanks{E-mail:koushik.physics@gmail.com}\ ,
Anirban Roy$^{3,4,5}$\thanks{E-mail: aroy@sissa.it}\ , 
Ruchika$^{6}$\thanks{E-mail:ruchika@ctp-jamia.res.in}\ ,
Anjan A. Sen$^{6}$\thanks{E-mail:aasen@jmi.ac.in}\ , \\ \vskip -3mm
M.M. Sheikh-Jabbari$^{7,8}$\thanks{E-mail:jabbari@theory.ipm.ac.ir }}\\
\vskip 5mm
$^{1}$ Department of Physical Sciences, Indian Institute of Science Education and Research Kolkata, Mohanpur, West Bengal 741246, India\\
$^{2}$ Theory Division, Saha Institute of Nuclear Physics, HBNI,\\ 1/AF Salt Lake, Kolkata-700064, India\\
$^{3}$ Department of Astronomy, Cornell University, Ithaca, NY 14853, USA\\
$^{4}$SISSA, Via Bonomea 265, 34136, Trieste, Italy\\
$^{5}$ Institute for Fundamental Physics of the Universe (IFPU), Via Beirut 2, 34014 Trieste, Italy\\
$^{6}$Centre for Theoretical Physics, Jamia Millia Islamia, New Delhi-110025, India.\\
$^{7}$School of physics, Inst. for research in fundamental sciences (IPM),\\ P.O.Box 19395-5531, Tehran, Iran\\
$^{8}$ The Abdus Salam ICTP, Strada Costiera 11,  34151, Trieste, Italy\\ }

\maketitle
%\date{\today}

\begin{abstract} 
\noindent
Assuming that the Universe at higher redshifts ($z \sim 4$ and beyond) is consistent with \lcdm model as constrained by the Planck measurements, we reanalyze the low redshift cosmological data to reconstruct the  Hubble parameter as a function of redshift. This enables us to address the $H_0$ and other tensions between low $z$ observations and high $z$ Planck measurement from CMB. From the reconstructed $H(z)$, we compute the energy density for the ``dark energy'' sector of the Universe as a function of redshift {without assuming a specific model for dark energy}. We  find that the dark energy density has a minimum for certain redshift range and that the value of dark energy at this minimum $\rde^{\text{min}}$ is negative. This behavior can most simply be described by a {negative cosmological constant} plus an evolving dark energy component. We discuss possible theoretical and observational implications of such a scenario.
\end{abstract}
%\pacs{98.80.Es, 95.36.+x , 98.80.-k, 98.65.Dx}
%\begin{keywords}

\section{Introduction}
{Thanks to various sets of cosmological data, we can now talk about ``{\it the standard model of cosmology}'', the $\Lambda$CDM Universe \cite{Ade:2015xua, Ade:2015rim}. It provides the simplest paradigm that fits remarkably well to most of the current cosmological observations.  As the precision of the low redshift data increases, there are however emerging tensions in \lcdm model which is otherwise consistent with high redshift CMB observations by Planck. The major tension is between the model independent measurement of $H_{0}$ parameter ({$\sim 73 ~\text{km/s/Mpc}$}) by SH0ES collaboration \cite{Riess:2016jrr, Riess:2017lxs, Riess:2018byc} and that by Planck assuming \lcdm model \cite{Ade:2015xua, Ade:2015rim, Akrami:2018vks,Aghanim:2018eyx}. The latest Planck-2018 data shows, {$H_0 = 67.8 \pm 0.9~ \text{km/s/Mpc}$} \cite{Akrami:2018vks} and  this is at tension over $3.5\sigma$ with the SH0ES measurement which is $H_0 = 73.52 \pm 1.62~\text{km/s/Mpc}$.  A similar mild inconsistency in $H_{0}$ for $\Lambda$CDM model is also observed by Strong Lensing experiments like H0LiCow using time delay measurements \cite{Bonvin:2016crt} which measured $H_0 = 71.9^{+2.4}_{-3.0}~\text{km/s/Mpc}$ for \lcdm Universe.  Moreover the BOSS survey for baryon acoustic oscillations measurements using Lyman-$\alpha$ forest \cite{Delubac:2014aqe} has also measured the expansion rate of the Universe at $z=2.34$. This measured expansion rate of the Universe at $z=2.34$ is also at tension over $2\sigma$ with Planck result for \lcdm \cite{Sahni:2014ooa}. The important consequences of these tensions are the prediction of the dark energy density evolution with time\footnote{{That evolving and dynamical dark energy is necessitated by the data has also been discussed in \cite{Sola:2018sjf, Sola:2016ecz,Ryan:2018aif, Ooba:2018dzf, DiValentino:2017zyq, Rivera:2016zzr, Zhao:2017cud, Zhang:2017idq}.}} {and more importantly for us,}  the possibility of having  negative dark energy density at higher redshifts as discussed by \cite{ Delubac:2014aqe, Sahni:2014ooa, Poulin:2018zxs}. Similar conclusion has also been obtained recently by \cite{Wang:2018fng} using a dark energy model independent approach. In their study, Wang et al. attributed the evolution of dark energy density as well as its negative values at high redshifts to non-minimally coupled scalar field theory like Brans-Dicke theory. Interestingly in all these studies, the dark energy density is not only negative at higher redshifts but it is unbounded from below. This poses a serious problem as in a spatially flat universe, with such a negative dark energy density, the matter energy density parameter will be more than one and grow quickly without any upper bound at higher redshifts. This will have catastrophic effect on the structure formation scenario.This surely needs to be resolved.

%%
%%Earlier Sahni et al. \cite{sahni} have also confirmed similar tension between $H$ at $z\sim 2.4$ as measured by Lyman-$\alpha$ BAO observations and the same obtained by Planck for \lcdm model. Using a combination of cosmological observations, Zhao et al. \cite{gongbo} have obtained a model independent constraint on the equation of state (EoS) of dark energy which is not consistent with $w=-1$. Recently, Wang et al. \cite{Wang:2018}  have studied model independent constraint on the evolution of the dark energy density and confirmed that with a evolving dark energy, the above mentioned inconsistency in $H_{0}$ is reduced. Moreover, \textcolor{red}{and more importantly for us,} they showed that the dark energy density is likely to fall below zero in the past. They observed that this is primarily due to the BAO measurement with Lyman-$\alpha$ forest at $z\sim 2.4$ \cite{evslin,sahni} and attributed this negative $\rho_{DE}$ to non-minimally coupled scalar field theory like Brans-Dicke theory.
%%

In this work, we revisit the process of constraining the dark energy behavior using cosmological observations to find the likely sources of tensions between low and high redshifts cosmological observations. For this, }we should be careful that although Planck constrains the high redshift Universe  using CMB with unprecedented accuracy, it may not be sensitive enough to any new physics beyond \lcdm at low redshifts. Fitting the entire background evolution of the Universe from $z\sim1100$ till $z=0$ using $\Lambda$CDM in which the dark energy density is a redshift independent quantity, can be the source of aforementioned tensions.

{Therefore we reanalyze the low-redshift data involving background cosmology assuming that for a certain higher redshift $z=z_{\text{match}}$ and beyond, one recovers the background evolution as constrained by Planck-2018 for \lcdm Universe. {For} this, we assume that $z_{\text{match}}$ has to be larger than  $z\sim 2.4$, as {around this redshift}, we have the constraint on expansion rate of Universe from BOSS survey using Lyman-$\alpha$ forest. This takes care of any new physics for dark energy evolution (beyond \lcdm Universe) that is predicted by low-redshift observations for $z < z_{\text{match}}$ including the SH0ES measurement for $H_{0}$. 
{With these assumptions, we study the dark energy behaviour which is consistent with the low-redshift data for $z\leq z_{match}$ {while smoothly matching} the Planck-constrained $H(z)$ behaviour for $\Lambda$CDM model for $z>z_{match}$}. In this process, we do not assume any particular dark energy model. We directly reconstruct the Hubble parameter $H(z)$ as a function of redshift using cosmographic approach. Subsequently, we reconstruct the dark energy density evolution} {in a  ``dark energy relaxed XCDM framework,'' in which we do not assume any specific dark energy model,}\footnote{{We note that similar reconstruction of dark energy properties have been extensively analyzed  in the literature, e.g. see \cite{Shafieloo:2007cs, Seikel:2012uu, Gerardi:2019obr, Wang:2019ufm}. Our data analysis method and the data sets we consider here is different than these other works.}} {while still assume  the Universe is spatially flat and contains pressure-less matter (including baryons and dark matter) plus another unknown component. This extra unknown component can be due to dark energy or it may arise as an effective dark energy component due to any sort of modification of gravity at large cosmological scales. We also  ignore the contribution from radiation energy density as it is negligible compared to matter or dark energy at late times.}

%%
%Therefore we reanalyze the low redshift data for background cosmology while using the Planck-2018 constraints on background evolution (for \lcdm model) \cite{ade1} as an input at redshift around $z=z_{match} $. This will lead to Hubble parameter as a function of $z$ that is consistent with low redshift data for $z\leq z_{match}$ (including that from R18 \cite{R18}) and for $z>z_{match}$, it is also consistent with $H(z)$ for \lcdm Universe as constrained by Planck. {We reconstruct $H(z)$ evolution from the data in a ``dark energy relaxed \lcdm framework'', in which we do not assume any specific dark energy model. 
%We do not assume any particular dark energy model for this purpose; rather we directly constrain the $H(z)$ evolution in a model independent approach. 
%Subsequently, we reconstruct the dark energy evolution assuming that the Universe is spatially flat and contains pressure-less matter (including baryons and dark matter) and dark energy with a negative pressure.}

Our {dark energy model-independent} analysis within the above set of working assumptions and framework reveals that $\rde(z)$ has two specific features: it has a minimum and a phantom crossing at  $z=z_{\text{min}}$. Moreover, the value of $\rde$ at $z=z_{\text{min}}$ is negative. The simplest explanation for $\rde^{\text{min}} < 0$, is the existence of a {\it negative Cosmological Constant}.

\section{Data sets and their analysis} There are different approaches to constrain the behavior of dark energy without assuming any particular dark energy model. Here we take the cosmographic approach that directly constrains  different kinematical quantities related to the  background evolution. The Taylor expansion of $H(z)$ for this purpose does not work for $z>1$ and hence we use the Pade Approximation that improves the convergence at higher redshifts \cite{Saini:1999ba,Gruber:2013wua, Wei:2013jya, Rezaei:2017yyj, Mehrabi:2018oke, Benetti:2019gmo}. 

For our purpose we model the Hubble parameter $H(z)$ using Pade approximation $P_{2,2}$:
\begin{equation}\label{H-Pade}
%E(z) = \frac{H(z)}{H_{0}} = \frac{1+P_{1}z+P_{2}z^{2}}{1+Q_{1}z+Q_{2}z^{2}},\nonumber\\
H(z) = H_{0}  \frac{1+P_{1}z+P_{2}z^{2}}{1+Q_{1}z+Q_{2}z^{2}}.
\end{equation}
All the constant parameters $P_{1}, P_{2}, Q_{1}, Q_{2}$  appearing in the above expression, can be written in terms of the kinematic quantities like, $q, j, s, l$ (deceleration, jerk, snap, and lerk)  (see  \cite{Capozziello:2018jya} for details). Note that for \lcdm model, the jerk parameter  $j=1$ for all redshifts and any deviation from $j=1$, confirms a non-$\Lambda$CDM behavior.

{In the following two subsections we reconstruct $H(z)$ in two cases to highlight the effects of inclusion of {constraint on background expansion at high redshifts from Planck measurements of CMB anisotropy}: In section \ref{sec:2.1} we only consider the low redshift data samples mentioned below using Pade approximant. In section \ref{sec:2.2} we reanalyze the low redshift data requiring $H(z)$ passes through three data points at $z_{\text{match}}, z_{\text{match}}\pm 1$. {The former may then be used to compare with  results of \cite{Zhao:2017cud, Wang:2018fng}. This analysis may be viewed as a check for the Pade parametrization used.}}

\subsection{Low redshift data only}\label{sec:2.1}

{
To constrain the background cosmology, we use the following set of low-redshift observational data:
\begin{itemize}
\item
BAO measurements from different surveys including eBoss quasar clustering and  Lyman-$\alpha$ forest samples (see \cite{Evslin:2017qdn} and references therein);\vskip 2mm
\item
latest Pantheon data for SNIa \cite{Riess:2017lxs, Gomez-Valent:2018hwc};\vskip 2mm
\item
the OHD data for Hubble parameter at different redshifts \cite{Pinho:2018unz};\vskip 2mm
\item
strong lensing time-delay measurements by H0LiCOW experiment \cite{Bonvin:2016crt};\vskip 2mm
\item angular diameter distances measured using water megamasers under the Megamaser Cosmology Project \cite{Evslin:2017qdn, Reid:2012hm, Kuo:2012hg, Gao:2015tqd};\vskip 2mm
\item
finally the measurement of $H_{0}$ by \cite{Riess:2016jrr}[R16].
\end{itemize}}
For details about all the data used in this analysis, please refer to \cite{Capozziello:2018jya}. The independent parameters for the data analysis are $H_{0}, q_{0}, j_{0}, s_{0}$, $l_{0}$  and $r_{d}$, the sound horizon at drag epoch. Here the subscript $``0"$ means the value at present ($z=0$).

\begin{figure}
\begin{center} 
%\resizebox{200pt}{200pt}{\includegraphics{Hwithoutplanck.pdf}}
\resizebox{200pt}{200pt}{\includegraphics{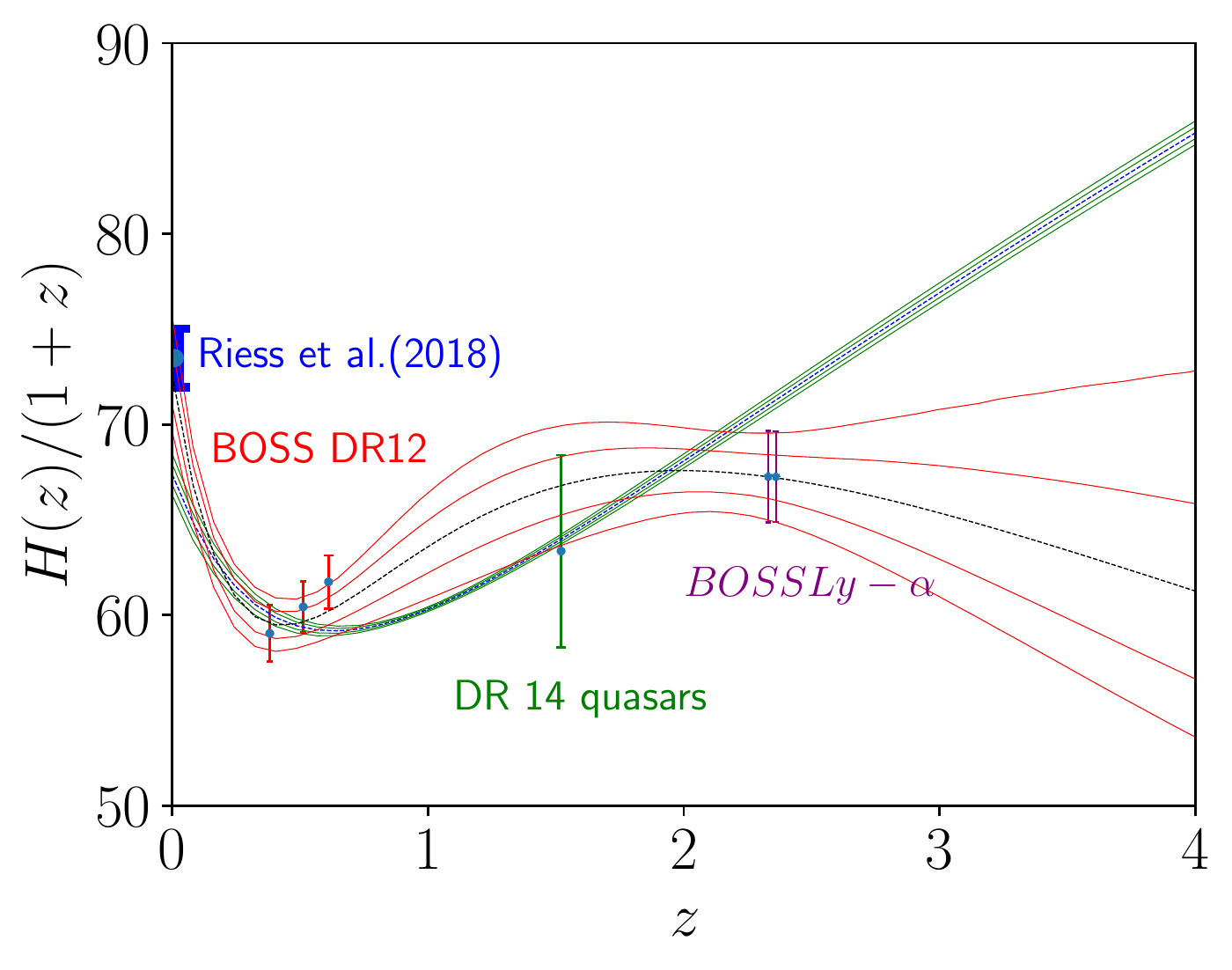}}
\resizebox{200pt}{200pt}{\includegraphics{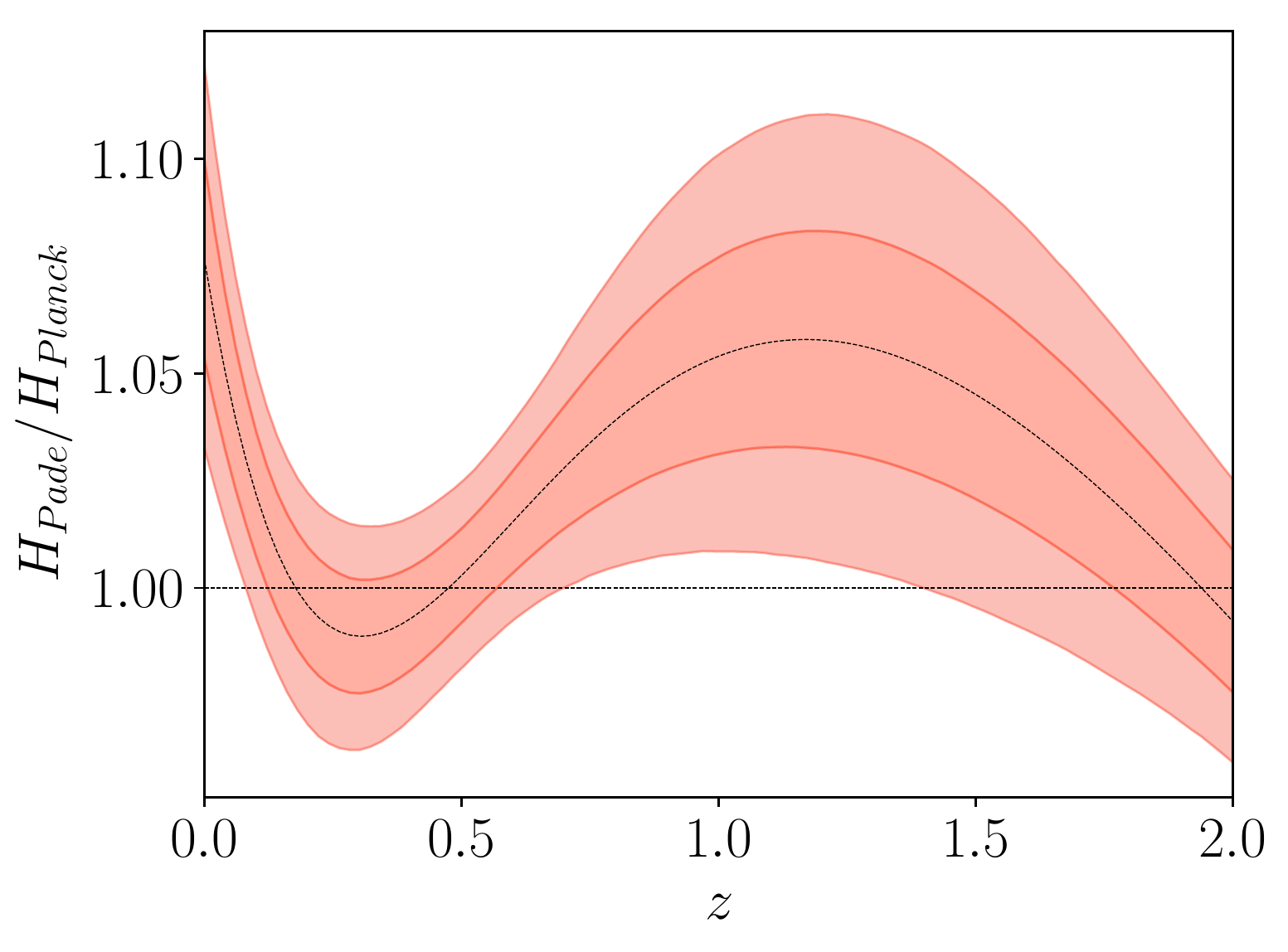}} 
\resizebox{200pt}{200pt}{\includegraphics{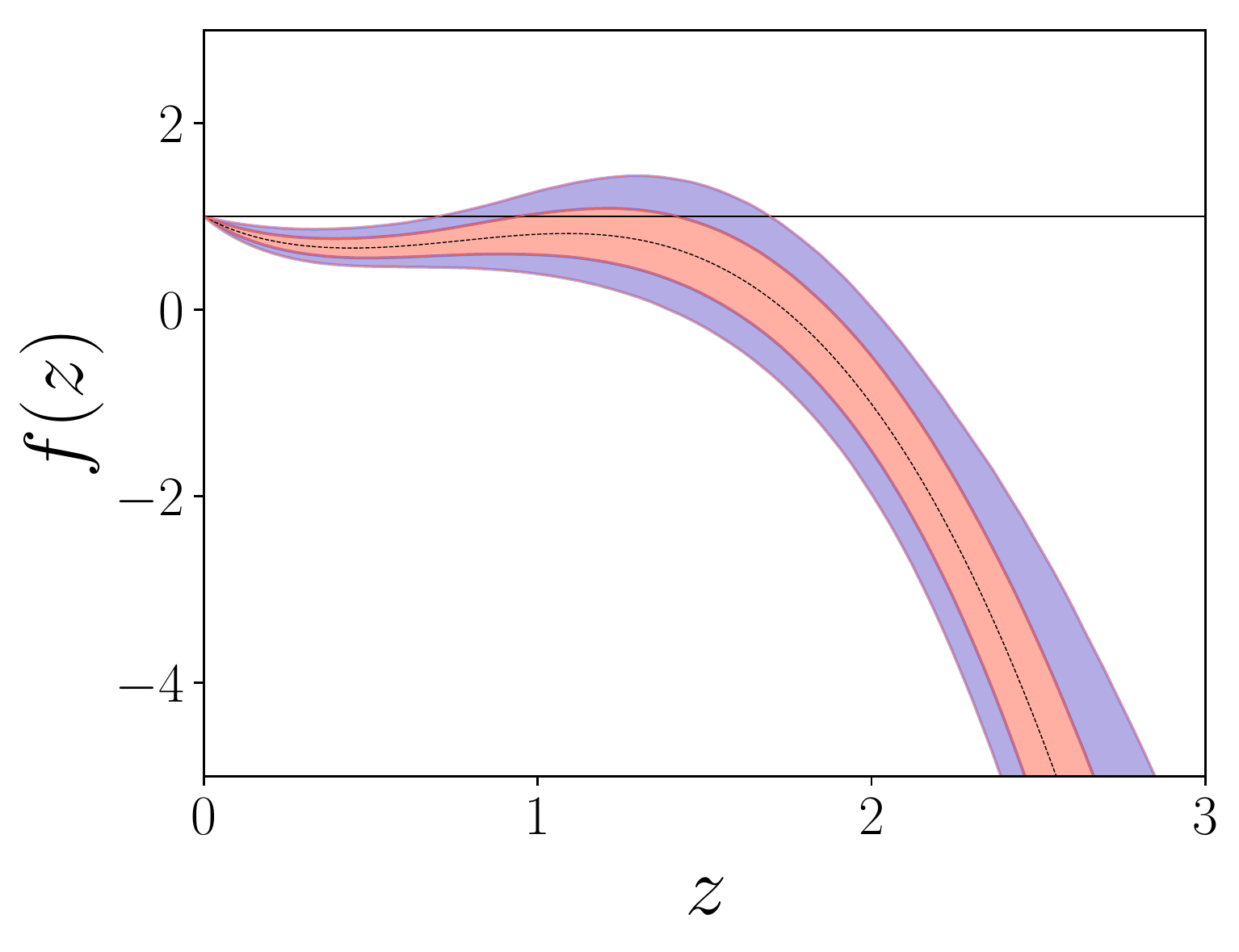}}
\end{center}
\caption{{Low data redshift analysis without inclusion of high redshift CMB data from Planck.} Left-Top: Reconstructed Hubble parameter $H(z)$  from various low-redshift data sets. The black dashed line is for best fit values whereas the inner and outer lines denote $1\sigma$ and $2\sigma$ contours. The thin shaded region is reconstructed $H(z)$ behaviour from Planck-2018 measurements. Right-Top:  Comparison of our reconstructed $H(z)$ using Pade approximant and the best fit $H(z)$ for $\Lambda$CDM by Planck-2018. Bottom: Reconstructed dark energy density as a function of redshift. The horizontal line $f(z)=1$ is for $\Lambda$CDM.}\label{fig:hwoutpl}
\end{figure}

{ The reconstructed $H(z)$ is shown in Fig.\ref{fig:hwoutpl} (Left-Top plot). {Note that the high uncertainty beyond $z\sim 2.4$ is due to unavailability of low redshift data beyond $z\sim2.4$}. As one can see (the Right-Top plot), the reconstructed $H(z)$ from low-redshift observations is at large tension with the same, reconstructed from Planck-2018 data. Also at low redshifts ($z \leq 2$), there is  oscillations in $H(z)$ behaviour around the best fit $H(z)$ behaviour for $\Lambda$CDM as measured by Planck-2018. Such oscillation in $H(z)$ has already been reported in  \cite{Wang:2018fng} through a model independent study. In this work, we obtain exactly the same behaviour in a completely different approach using Pade approximation for cosmography.}

Once we constrain the $H(z)$ behaviour using available low-redshift data, we can further constrain the redshift evolution of any exotic component which one needs to add to the matter contribution. This extra component (we denote it as $\rde$) can be due to dark energy or due to any modification of gravity on large cosmological scales that gives rise to some effective dark energy component. Following \cite{Wang:2018fng}, we write:\footnote{{In principle, one could have included a curvature term, $\rho_C (1+z)^2$, especially in light of recent claims in \cite{DiValentino:2019qzk}. We intend to study such a possibility in upcoming work. }}
\begin{equation}\label{H2}
3 H^{2}(z) = \rho_{m} + \rde = \rho_{m 0}  (1+z)^3 + \rho_{_{\text{DE}0}} f(z).
\end{equation}
Here we set $8\pi G = 1$ and superscript ``0" indicates values at present $z=0$ and subscript ``DE'' stands for any dark component (either actual or effective) that yields late time acceleration. The $f(z)$ (a dimensionless quantity)  specifies the allowed redshift evolution of $\rde$. Assuming a flat Universe,  $\Omega_{m} + \Omega_{\text{DE}} = 1$, equation (2)  can be rewritten as
\begin{equation}
\frac{H^{2}}{H_{0}^{2}} = \Omega^{(0)}_{m} (1+z)^3 + \Omega^{(0)}_{\text{DE}} f(z) 
= \Omega^{(0)}_{m}(1+z)^3 + (1-\Omega^{(0)}_{m}) f(z).
\end{equation}
With this, one can constrain the redshift evolution for $\rde(z)$ using the behaviour for $H(z)$  shown in Fig.\ref{fig:hwoutpl}. For this, one needs to assume the value of $\Omega^{(0)}_{m}$ and for our purpose, we assume $\Omega^{(0)}_{m} = 0.315$ which is best fit value for Planck-2018 for $\Lambda$CDM and this value is not very sensitive to dark energy behaviour. Using this, one can now reconstruct the $f(z)$ behaviour from low-redshift observation which is shown in Fig. \ref{fig:hwoutpl} (Bottom plot). As one can see, the $f(z)$ clearly shows oscillations around the $f(z)=1$ $\Lambda$CDM behaviour for low redshifts. This confirms the phantom-non phantom crossing. Also for larger redshifts, $f(z) <0$ and is also unbounded from below. These behaviours of $f(z)$ are in complete agreement with the results obtained by \cite{Wang:2018fng}, confirming our methodology of using Pade approximations.

%\newpage
\subsection{Low redshift data plus inclusion of CMB Planck data points for \textit{H(z)}}\label{sec:2.2}

{Once we confirm the results obtained by \cite{Wang:2018fng}  as well as other authors \cite{Sahni:2014ooa, Delubac:2014aqe, Poulin:2018zxs} with different approach, our next goal is to solve the problem of $\rde$ being unbounded from below for negative values for larger redshifts. To address this problem, we assume that $H(z)$ as constrained by the low redshift data, should also fit $H(z)$ as measured by Planck for \lcdm at some higher redshift $z=z_{\text{match}}$ and beyond. }{This assumption is well justified noting that the difference between our XCDM model and \lcdm is in their dark energy sector and that the dark energy effects becomes relevant to the evolution of the Universe {only at lower redshifts}. 
{Also as discussed in the Introduction, to take into account any new physics in the dark energy sector that may arise due to the BAO observations at $z=2.4$ using Lyman-$\alpha$ forest,}\footnote{{We have repeated the analysis with the model independent diameter distance measurements \cite{Carvalho:2015ica, Alcaniz:2016ryy, Carvalho:2017tuu} 
and find that they do not change our results.}}   $z_{\text{match}}\gtrsim 4$ should be a reasonable choice. Of course, as we will comment in the discussion part, in a full and complete analysis, $z_{\text{match}}$ should also come out from the analysis itself. Here we only intend to present a ``proof of concept'' analysis.} 

\begin{figure}%[t]
\begin{center} 
%\resizebox{200pt}{200pt}{\includegraphics{pl1.pdf}} 
%\hspace{2mm} \resizebox{200pt}{200pt}{\includegraphics{pl2.pdf}}
\resizebox{280pt}{220pt}{\includegraphics{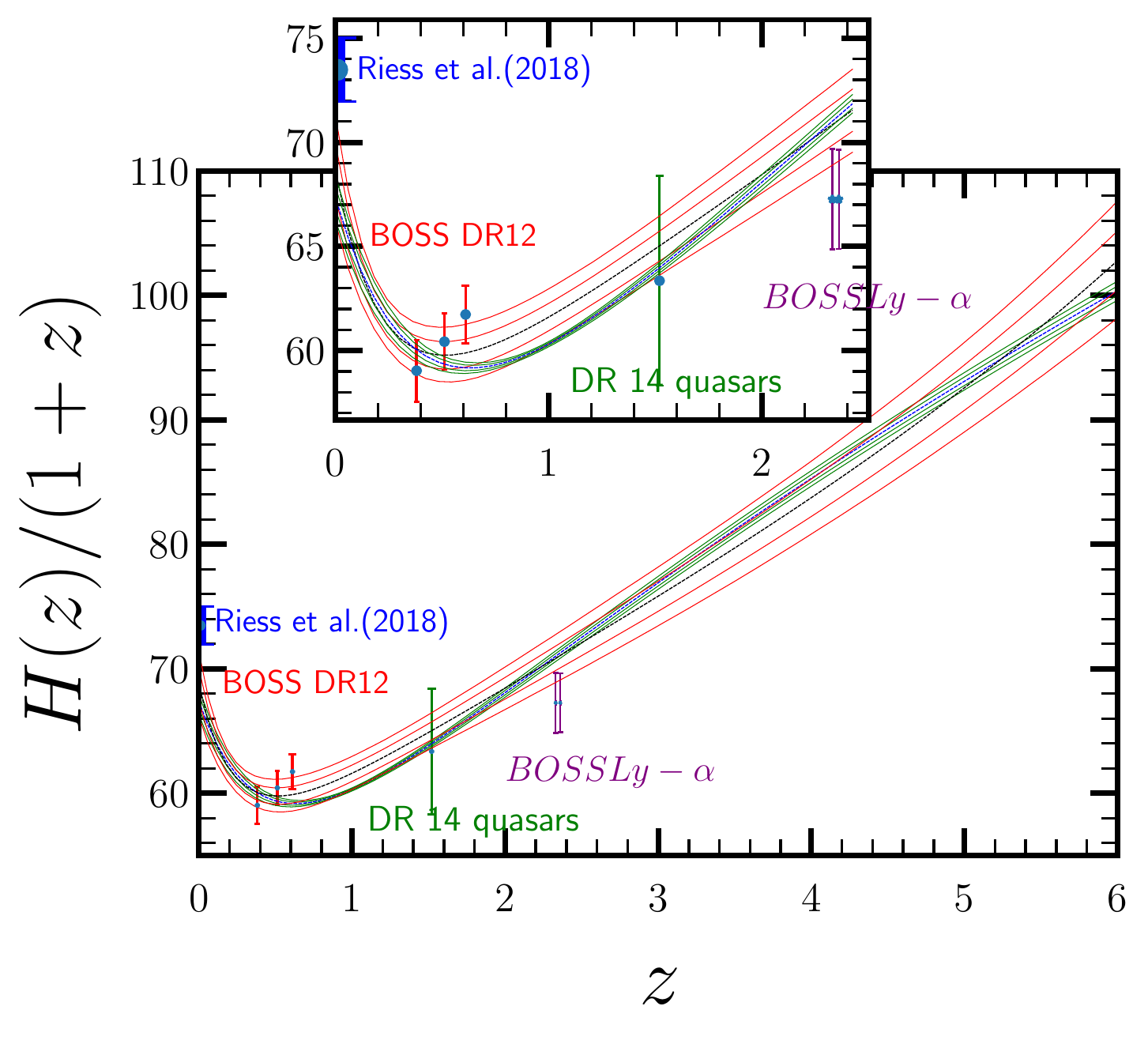}} \\
 \resizebox{280pt}{220pt}{\includegraphics{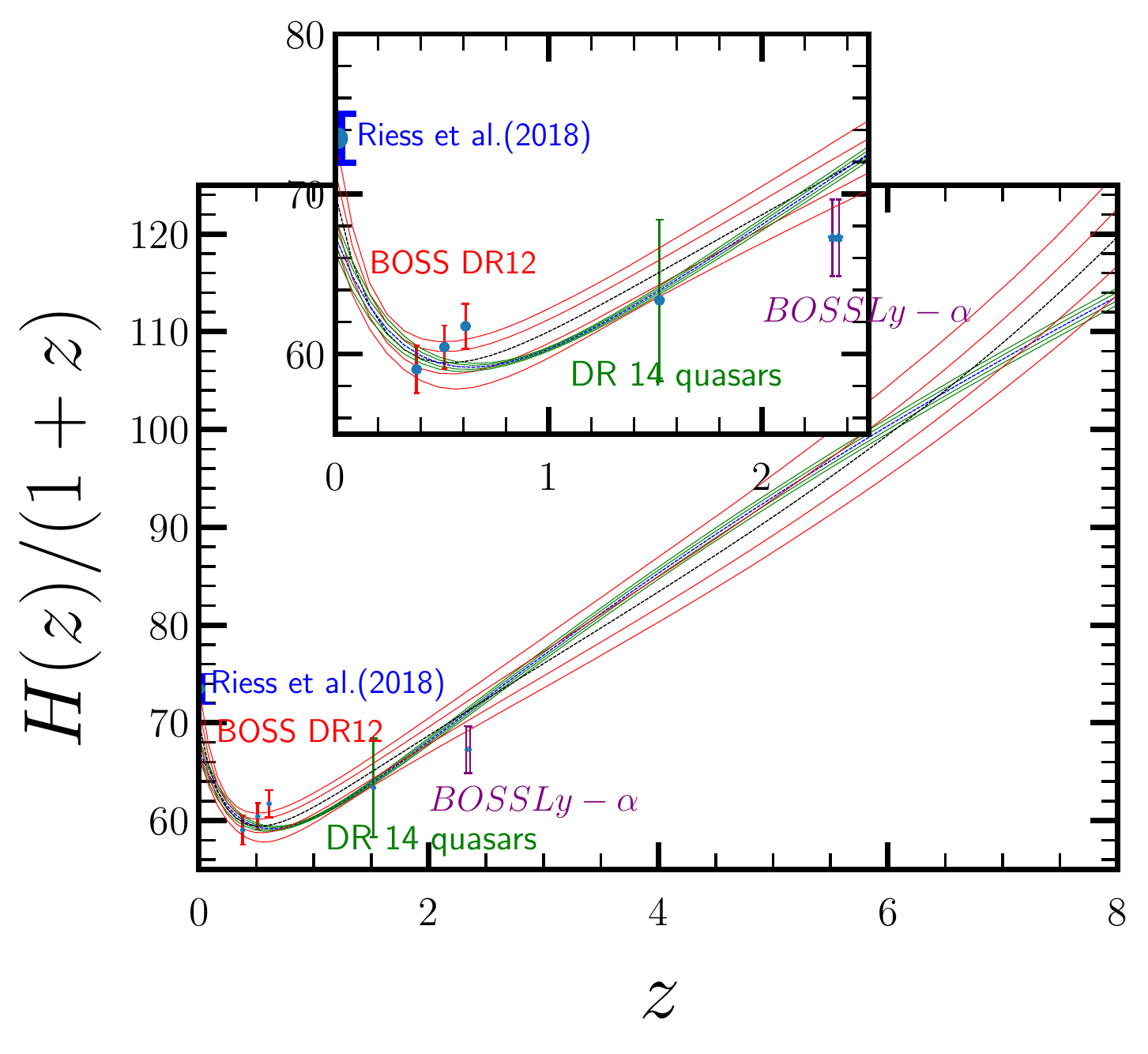}}
\end{center}
\caption{Reconstructed Hubble parameter $H(z)$ behavior {employing both low redshift and CMB Planck data sets, see the text for details}. The left one is based on Planck $H(z)$ data at $z=4,5,6$ and right one is for $z=6,7,8$.  The dashed line is for mean and inner and outer regions are for $68\%$ and $95\%$ confidence regions. The thin shaded region is reconstructed $H(z)$ behaviour from Planck-2018 measurements. }\label{fig:hqplot}
\end{figure}

To this end, we generate $H(z)$ data points for higher redshifts using Planck sample chains and use those data for the fitting. We use two sets of data for this purpose: in one case we generate data for $H(z)$ using Planck \lcdm chains for redshifts $z=4,5,6$ [hereafter we refer this case as PL1] {assuming $z_{match} \sim 4$} and in another case we do the same for $z=6,7,8$ [hereafter we refer his case PL2] {assuming  $z_{match} \sim 6$}. In both cases, we use sample chains for Planck+WP+highL+lensing-2015 for baseline \lcdm model \cite{plchain}.

The reconstructed Hubble parameter, as a function of $z$, is shown in Fig.\ref{fig:hqplot}. As one can see, the reconstructed $H(z)$ fits the $H_{0}$ data from R16 as well as $H(z)$ points as measured by Planck-2015 for \lcdm model \cite{Ade:2015xua} at higher redshifts. The constraints on deceleration and jerk parameters at $z=0$ are: ($q_{0} = -0.83^{+0.10}_{-0.10}, j_{0} = 3.93^{+0.53}_{-0.53}$) and ($q_{0} = -0.94^{+0.11}_{-0.10}, j_{0} = 4.31^{+0.51}_{-0.53}$) for PL1 and PL2 respectively. Clearly $j_{0} =1$ is ruled out with high confidence level confirming the inconsistency with \lcdm model at present. This is consistent with previous results by \cite{Zhao:2017cud} and \cite{Wang:2018fng}.

\begin{figure}
\begin{center} 
\resizebox{200pt}{170pt}{\includegraphics{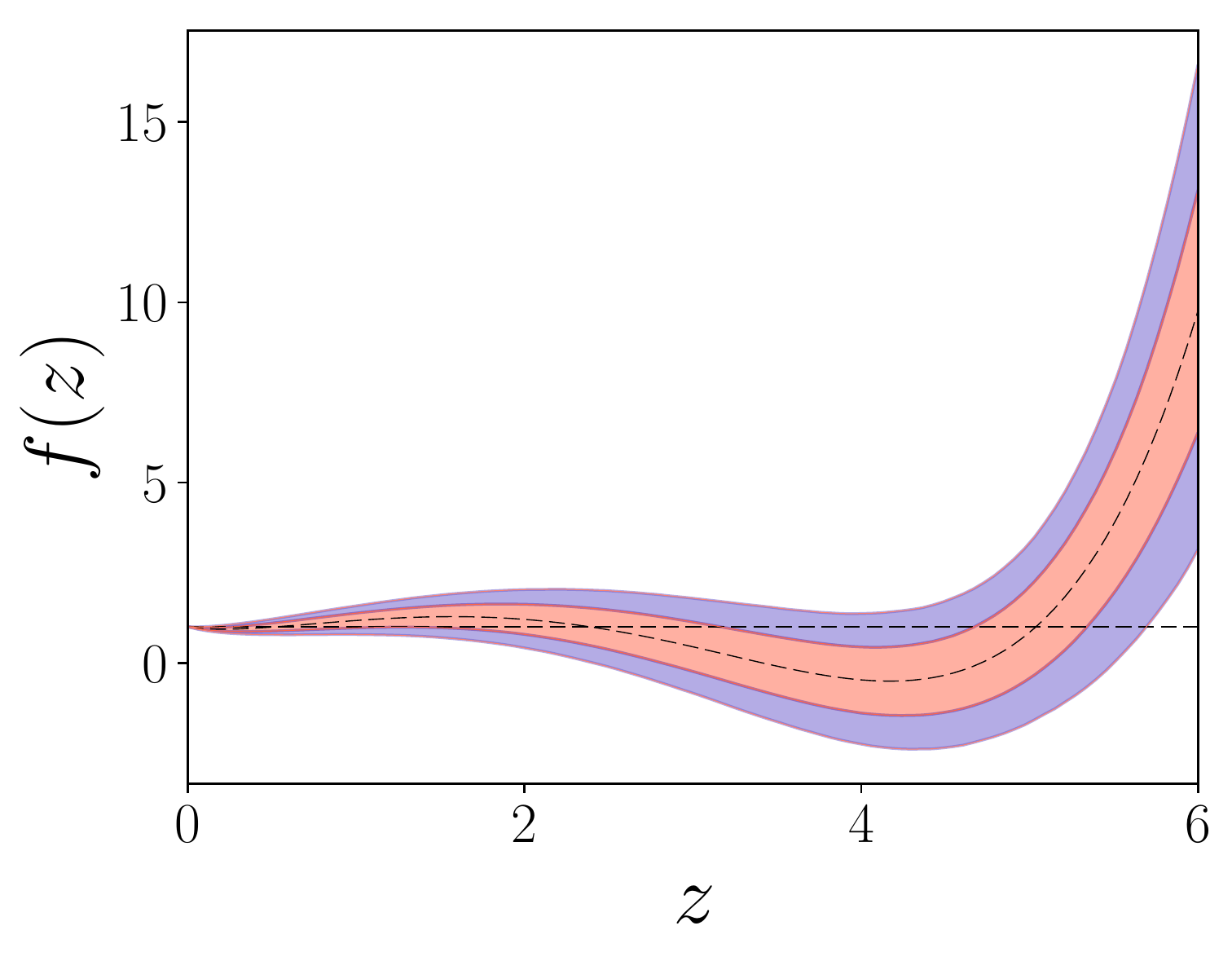}} 
\hspace{1mm} \resizebox{200pt}{170pt}{\includegraphics{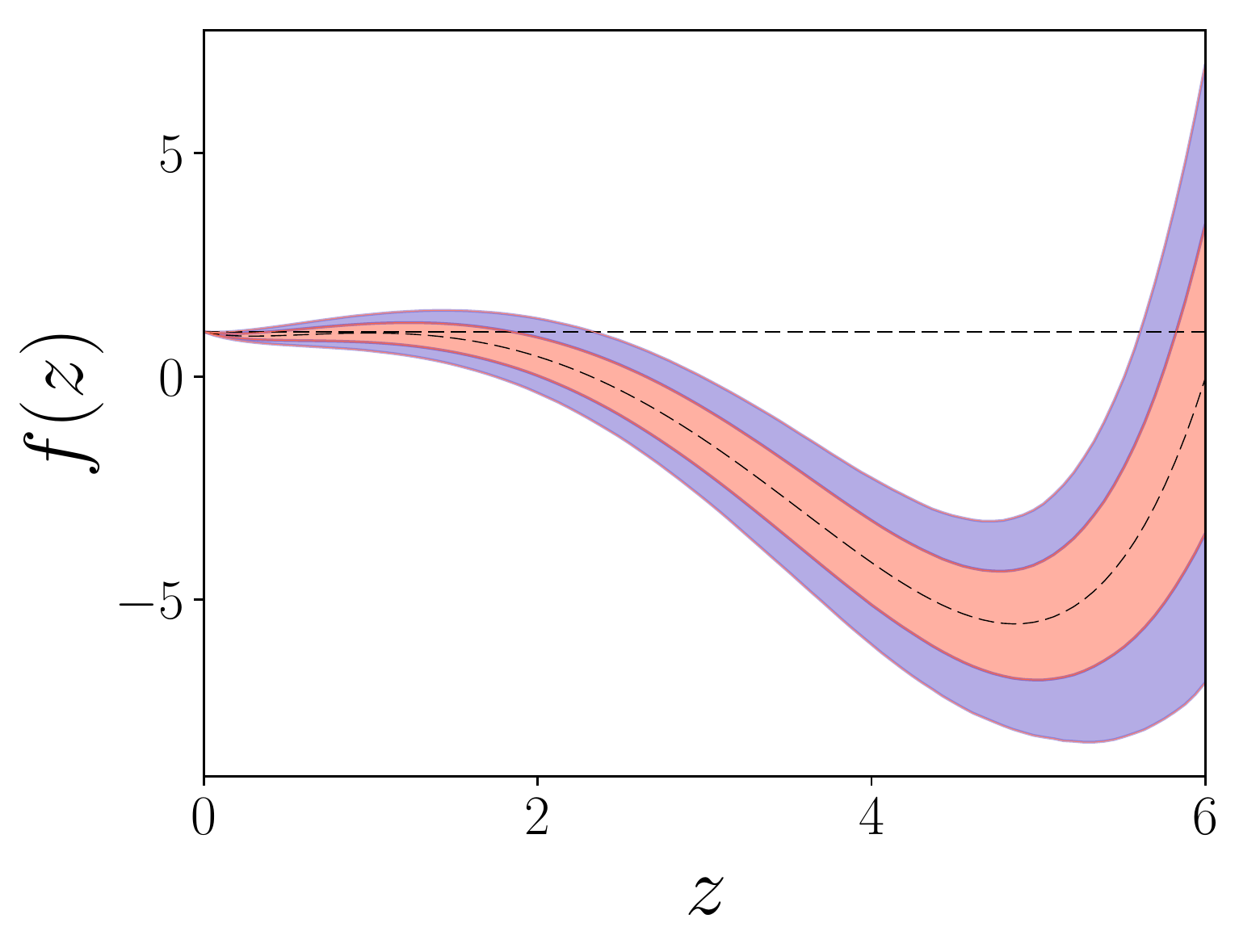}}\\
 \resizebox{200pt}{170pt}{\includegraphics{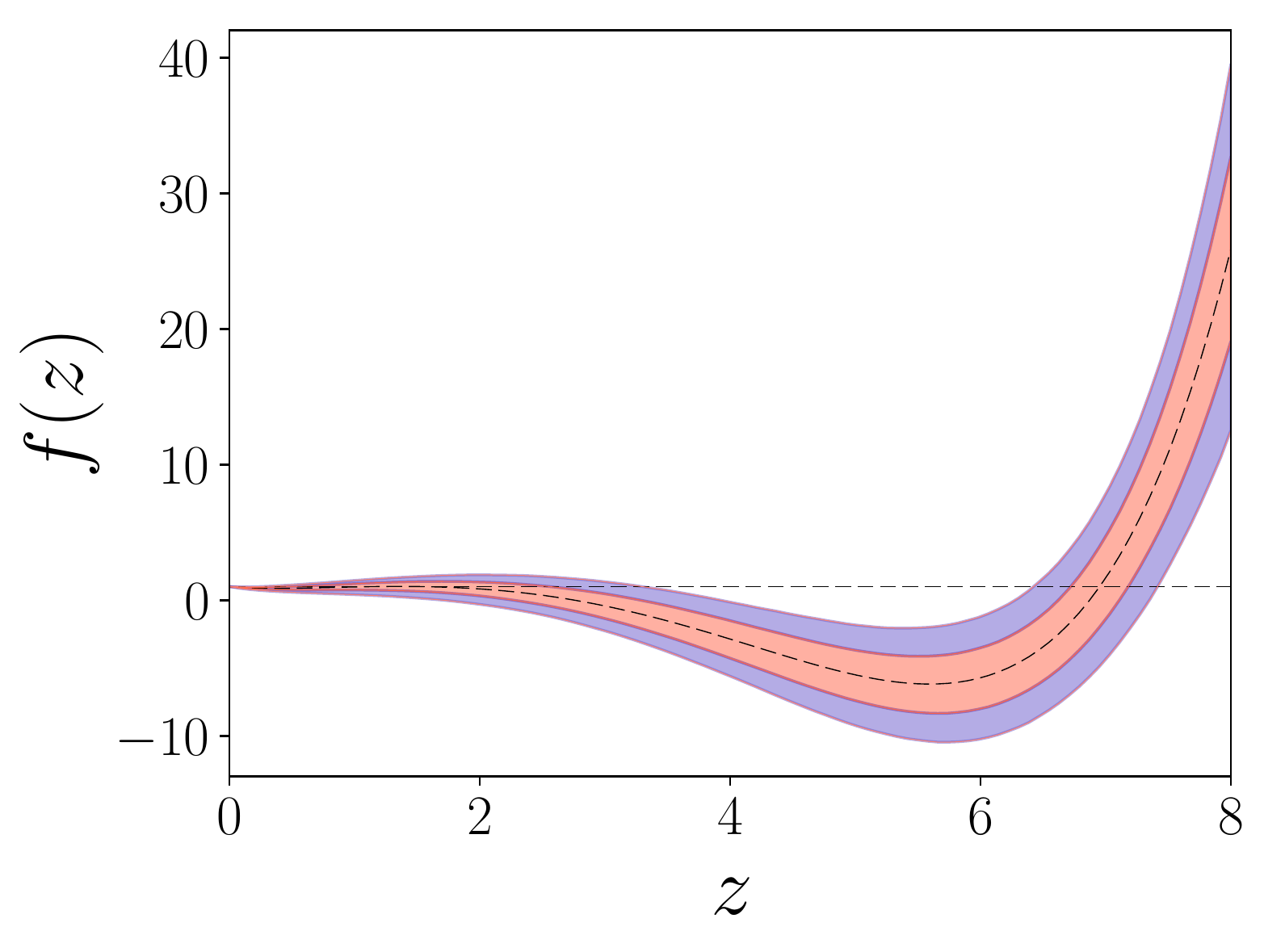}} 
\hspace{1mm} \resizebox{200pt}{170pt}{\includegraphics{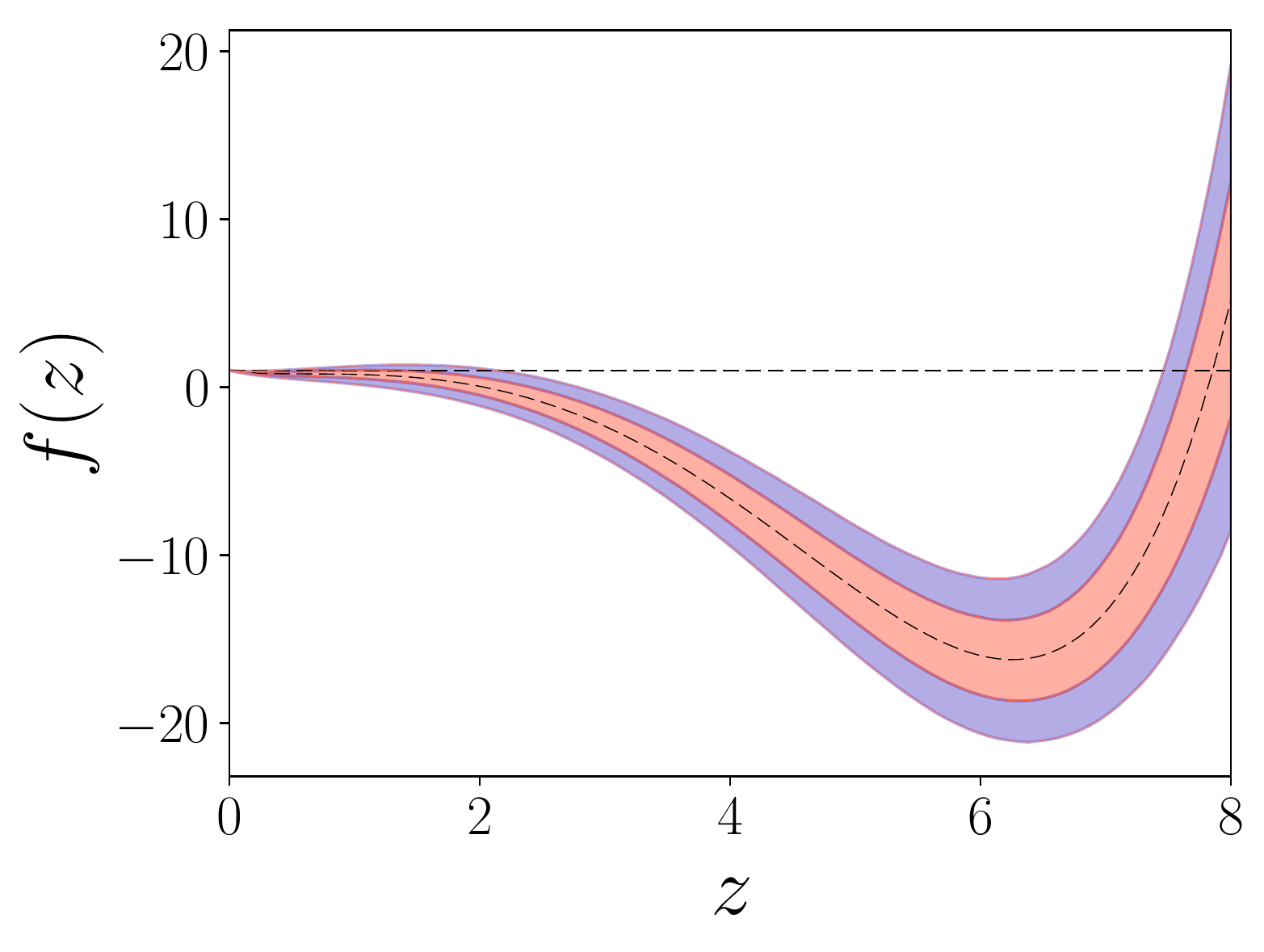}}
\end{center}
\caption{Reconstructed $f(z)$. The top ones for PL1 whereas the bottom ones are PL2. In each case, left ones are for $\Omega^{(0)}_{m} = 0.3$ and right ones are for $\Omega^{(0)}_{m} = 0.32$. The different regions and lines are same as in Fig.\ref{fig:hqplot}.}
\label{fig:fplot}
\end{figure}

Once we constrain the $H(z)$ using above data points, we again reconstruct the dark energy behavior through $f_{_{\text{DE}}}(z)$ as described equations (4) and (5). Here instead of choosing a particular value for $\Omega_{m0}$, we choose two different values, $0.3$ and $0.32$ respectively to see effect of $\Omega_{m0}$ on reconstructed $f(z)$. The result is shown in Fig \ref{fig:fplot}.

%\begin{figure}
%\begin{center} 
%\resizebox{150pt}{150pt}{\includegraphics{fz_using_padeorig_315_om.pdf}}
%\resizebox{200pt}{200pt}{\includegraphics{H_Planck2018.pdf}} 
%\end{center}
%\caption{Reconstructed dark energy density as a function of redshift. The horizontal line is for $\Lambda$CDM. }\label{fig:fzpade}
%\end{figure}
%}

As one can see, $f(z)$ exhibits two generic features: (1) the overall shape for $f(z)$ is the same for different choices of $\Omega^{(0)}_{m}$ as well as the redshift range where we generate $H(z)$ data using Planck chains for $\Lambda$CDM. It has always a minimum at some $z=z_{\text{min}}$, and generically $f(z) < 0$ at this minimum; (2) $z_{\text{min}}$ and $f_{\text{min}}=f(z_{\text{min}})$ depend on the value of $\Omega^{(0)}_{m}$ and the range of redshifts where one takes the $H(z)$ data from Planck. For $\Omega^{(0)}_{m} > 0.29$,  there is always a negative minimum for $f(z)$, and for $H(z)$ data from the Planck at higher redshift range, the negative minimum for $f(z)$ exists with a greater confidence level. The presence of this negative minimum in $\rde(z)$ does not affect the background evolution. This can be seen from Fig.\ref{fig:hqplot} that there is no pathological behavior in $H(z)$ which is perfectly consistent with a late time accelerating Universe. 

%\section{Discussions on the generic results}

\section{Generic Results} $f(z)$ is essentially showing ``time dependence'' of the dark energy density $\rde(z)$, therefore the time derivative of $\rde$ vanishes at $\zm$:
$$
\dot{\rho}_{_{\text DE}}(z)|_{z=\zm} = 0\ \  \  \Longrightarrow \hspace{2mm} (\rde +\pde)_{z=\zm} = 0.
$$
That is, at the minimum, the $\rde$ behaves like a cosmological constant. Moreover, this minimum value being negative, can be simply modeled through a {\it negative cosmological constant} with $\Lambda=3H_0^2 (1-\Omega^{(0)}_{m})  f_{\text{min}}$. The most significant outcome of our analysis is that the low redshift data together with Planck constraint on $H(z)$ for \lcdm at higher redshifts allows for a {\it negative cosmological constant}.
\begin{figure}
\begin{center} 
\resizebox{250pt}{250pt}{\includegraphics{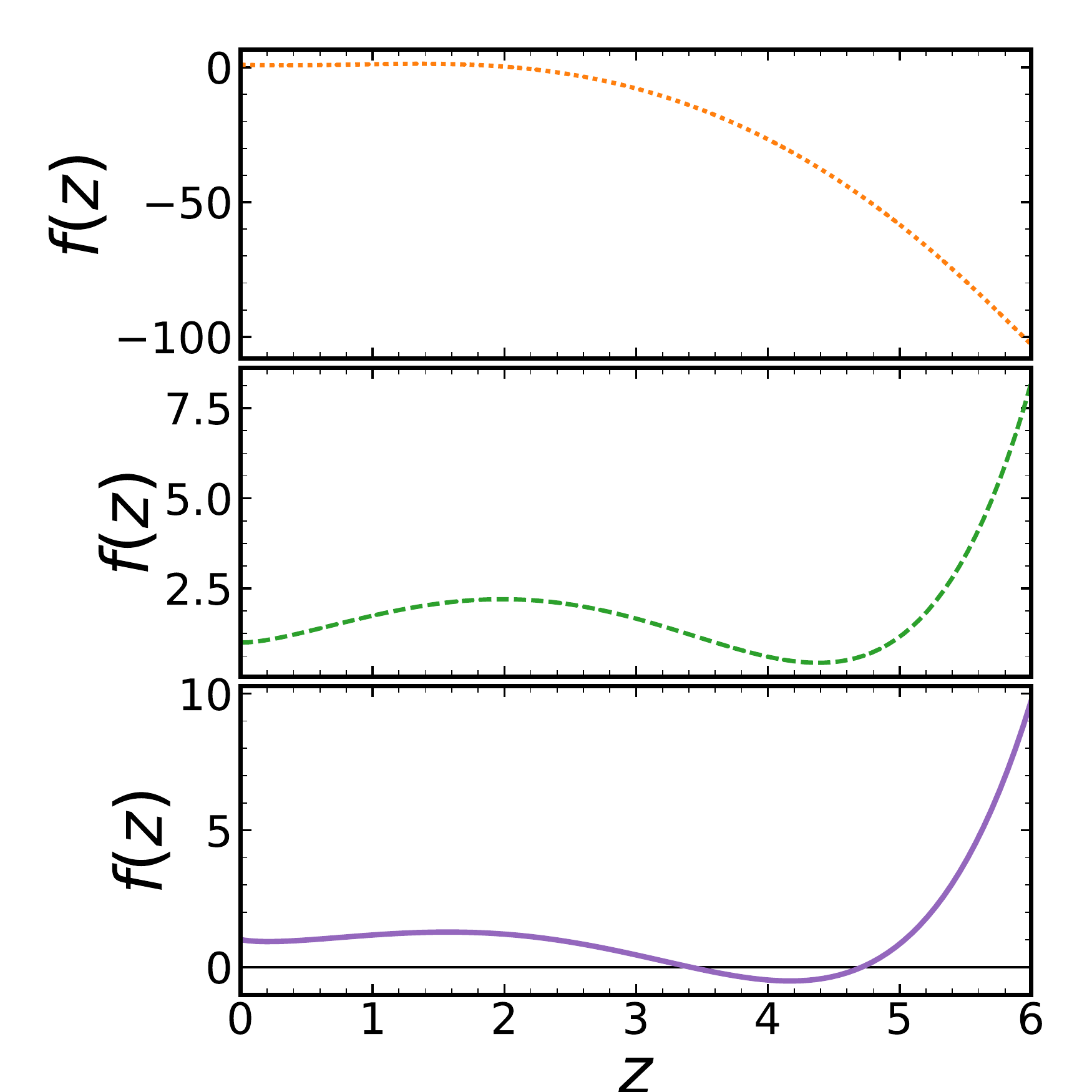}} 
\end{center}
\caption{ Reconstructed $f(z)$. The mean $f(z)$ is plotted from the MCMC chains. The top one is without $H_{0}$ data as well as without taking Planck points for $H(z)$ for higher redshifts. The middle one is without $H_{0}$ data but taking Planck points for $H(z)$ for higher redshifts. The bottom one is taking both the $H_{0}$ data and Planck Points for $H(z)$ for higher redshifts. The rest of the low redshift data as mentioned in the text are taken in all plots. Planck points for $H(z)$ are for PL1 and $\Omega^{(0)}_{m} =0.3$ is assumed.}
\label{fig:f1plot}
\end{figure}
As discussed in the Introduction, the negative $\rde$ without any minimum for higher redshifts has already been confirmed by \cite{Wang:2018fng} using low redshift data in a model independent way. The new feature in our analysis is the inclusion of the Planck's constraints on $H(z)$ for higher redshifts. This yields to a lower negative bound on $\rho_{DE}$ which can be associated with a \emph{negative Cosmological Constant}. To show these dependencies, in Fig. \ref{fig:f1plot} we plot the mean $f(z)$ behavior from our reconstruction for a specific case. The topmost  behavior is without including $H_{0}$ data from R16 and also without including $H(z)$ data generated using Planck \lcdm chains. The $f(z)$ clearly goes to negative values at higher redshifts as confirmed by  \cite{Wang:2018fng}, and this is mainly due to  Lyman-$\alpha$ measurement of BAO at $z\sim 2.4$ as pointed out in \cite{Wang:2018fng}, {see also \cite{Delubac:2014aqe, Poulin:2018zxs}}. Next we add the $H(z)$ data from Planck at higher redshift and this results a minimum in $f(z)$ which is positive. Finally, as we add the $H_{0}$ data from R16, the minimum shifts to a negative value. The inclusion of $H_{0}$ data from R16 enforces present Hubble parameter to be larger which can be compensated by lowering (even negative) dark energy density in the intermediate redshifts. Hence the Lyman-$\alpha$ data at $z\sim 2.4$, the $H_{0}$ data from R16 as well as the constraints from Planck on $H(z)$ at higher redshifts, all together play the crucial role for having a negative minimum for $\rde$ which may be modelled  by a \emph{negative Cosmological Constant}.

{The above conclusion can also be seen in this following way. Several previous analysis with low redshift data \cite{Wang:2018fng}, \cite{Delubac:2014aqe, Poulin:2018zxs} including our current analysis indicate that $f(z)$ becomes negative at some $z > 1$. At the same time, $H(z)$ should also go to matter domination at higher redshift. Then the only option for $f(z)$ is that it start with some positive value at $z=0$, starts decreasing for $z>0$, becomes negative at some point, attains a minimum with a negative value and then increases to $0$ for higher $z$, } {before matching to the Planck data at around $z_{\text{match}}.$} 
 
 %In Pade also, we have four parameters. 
%So we need to think of functions, which have a minimum at some $z=z_{min}$, which is +ve at z=0 and which vanishes at $z>=z_{match}$.
%
%Then we do the same fitting procedure. I am confident that we shall arrive at the same conclusions.
%
%Fixing the normalization of H(z) (to resolve H0-tension in favor of Riess et al), we need a four parameter family of parameterization, where three of them are essentially fixed by our inputs and the fourth is determined through the fit.  Of course, we are going to use this parametrization only for $0\leq z< z_{match}\sim 5--7$. 

 %\textcolor{red}{It is important to clarify that by ``model-independent" in the texts, we mean that we do not make a specific choice of the dark energy model. To be more precise, we keep all the ingredients of LCDM, but relax cosmological constant being the dark energy. From the LCDM and Planck best-fits, we use as an input the value of $\Omega_m$ and that $\Omega_m+\Omega{DE}=1$. We should of course make sure that the final results of our analysis do not upset the early time cosmology and the fit to CMB peaks and troughs (see later). 
%}

Moreover, for $z>\zm$, the $\rho_{_{\text{DE}}}$ decreases with time, whereas for  $z<\zm$, it increases with time, confirming the phantom crossing at $z=\zm$. The phantom region in this case cannot be mimicked by a minimally coupled scalar field rolling uphill the potential as discussed in \cite{Csaki:2004ha,Csaki:2005vq}. This is because for a minimally coupled scalar field $\phi$, at $\rho_{_{\text{DE}}}^{\text{min}}$, $\frac{d{\phi}}{dt} =0 $, which does not allow $\phi$ to roll uphill the potential.

\section{ Discussion and Outlook} 

Based on two requirements that $H_0$-tension is {ameliorated} (in favor of \cite{Riess:2016jrr}) and \lcdm is still the best fit for Planck data at higher redshifts, we reanalyzed the low redshift data. {If we consider just the low-redshift data, as we did in section \ref{sec:2.1}, we get $H_0=72.56 \pm 1.55$, which is perfectly consistent with \cite{Riess:2016jrr}. Adding the $H(z)$ data for higher redshift from Planck chains for \lcdm as we did in section \ref{sec:2.2}, we get the following results: For $z_{\text{match}}= 4$, $H_0=68.7 \pm 1.3$, which is $2.3\sigma$ tension with \cite{Riess:2018byc}. For
$z_{\text{match}}= 6$, $H_0=70.0 \pm 1.4$ which is at $1.65 \sigma$ tension with \cite{Riess:2018byc} and is at less than $2 \sigma$ tension with \cite{Riess:2019cxk}. The details of these analysis will appear in an upcoming paper. So the take away is: (1) adding the Planck constraint on background evolution at higher redshifts, the $H_0$ shifts towards lower values compared to only low-redshift analysis and (2)  adding the data point from $H(z)$ reconstructed by Planck chains, 
prefers  higher $z_{\text{match}}$. As we already pointed out  with higher $z_{\text{match}}$ the likelihood of having negative minimum in $f(z)$ is greater.
As we mention below,  the exact value of  $z_{\text{match}}$ should be obtained with a thorough analysis including low redshift and full CMB likelihood.
}

%in a model independent way 
We  reconstructed the $H(z)$ behavior {for $z\lesssim 8$ region}\footnote{{Note that we are using Pade $P_{2,2}$ parametrization only for $0<z_{\text{match}}\lesssim 8$ region and for higher $z$ {one should use higher order Pade parametrization for better fit to actual model}.}}  and subsequently the $\rde(z)$ behavior. We stated and discussed  three generic results of our %model independent
 analysis in the previous sections.  %Here we would like to add some further comments and things to be worked out and studied in future:
It is important to highlight the differences of our work from some recent works along this line. 
In the work by \cite{Zhao:2017cud}, the equation of state parameter for the dark energy was reconstructed with the assumption of $\rde > 0$, and also no interaction between matter and the dark energy, therefore, setting aside a very large class of dark energy models including modified gravity models. In a subsequent paper by \cite{Wang:2018fng}, $\rde$ was reconstructed directly, and it was found to be unbounded from below at higher redshifts. Almost similar inference was drawn in \cite{Poulin:2018zxs}. {In our analysis, we directly reconstruct the $H(z)$ from low redshift data using Pade approximation and get the similar result. Our analysis reproduces the result of \cite{Wang:2018fng}  that $\rde < 0$ at higher redshift. The fact that two different reconstructions yield similar results, supports  the validity of {our $H(z)$ reconstruction process which is based on Pade parametrization}.

\begin{comment}
{\color{green}In all the above cases, they incorporated the Planck constraint using acoustic scale, and we know that the major contribution to this scale, up to the surface of last scattering,  comes from low-redshift contribution. So in effect, they have not incorporated the high redshift constraint from the CMB data and the Planck.
As we have emphasized in the Introduction, {the data suggests we have a dynamical dark energy model with $z$-dependent equation of state.
%for the entire redshift range \footnote {We need to rephrase this sentence - Koushik will explain his confusion during discussions}. 
The Planck constraints the higher redshift universe with a greater accuracy, so we assume the high redshift $H(z)$ is given by what Planck measured for \lcdm\!\!. But in lower redshifts, where the effects of dark energy becomes important, there is new physics with non-LCDM, XCDM, behaviour, that solves low-redshift tensions including the $H_0$ one}}.({\bf this portion we can omit. It makes things a little confusing}).
\end{comment}

Subsequently we incorporated Planck constraint on background universe, directly through $H(z)$ as measured by Planck for \lcdm at intermediate redshifts. This makes sure that our reconstructed $H(z)$ from low redshift data is consistent with Planck measured $H(z)$ for \lcdm at intermediate redshifts.  This results in a minimum for $\rde$ and this minimum is negative. {The notable features of our approach for low-redshift data analysis are direct $H(z)$ reconstruction (using Pade approximant/parametrization) and imposing Planck constraint on $H(z)$ at intermediate redshifts.}
%and (iii) assuming two different behaviours for $H(z)$ at low and high redshifts.
}

Below, we add some further comments and things to be worked out and studied in future:
\begin{itemize}
\item[(I)]\textbf{{Vetting the working assumptions and robustness of our results:}} 
\vskip 2mm

This is probably the first study where we assume two different $H(z)$ behavior at low and high redshifts and match them around some $z=z_{\text{match}}$, {for which we chose some reasonable values. To justify this working assumption,} we have tested different $z_{\text{match}}$ and overall behavior of our results remain the same. This is also shown in Fig.\ref{fig:fplot}, where we plot $f(z)$ for two different choices of redshift range to generate $H(z)$ data from Planck \lcdm model. Of course, to get an actual estimate of $z_{\text{match}}$, one needs to do a full combined analysis using all the low redshift data together with Planck Likelihood assuming that for $z \leq z_{\text{match}}$, $H(z)$ is given by \eqref{H-Pade} and for $ z> z_{\text{match}}$, $H(z)$ is given by \lcdm model and get a estimate of $z_{\text{match}}$. This is beyond the scope of present study and will be reported in a separate publication.
\vskip 1.5mm

Our other working assumption was the use of Pade Approximation.  Pade approximation has been extensively used earlier for low-redshift reconstruction purposes, more detailed references can be found in \cite{Saini:1999ba} and more recent usage of Pade parametrization in \cite{Sahni:2014ooa, Aviles:2014rma, Gruber:2013wua, Wei:2013jya, Rezaei:2017yyj,  Mehrabi:2018oke} . Probably the first work on reconstruction of quintessence potential using Type-Ia Supernova data used the rational function like Pade Approximation to model the Luminosity Distance \cite{Saini:1999ba}. In our case, we use the same for the Hubble Parameter with larger set of observational data. We emphasis that use of Pade Approximation to reconstruct of Hubble parameter does not bias the final result. The result that dark energy density can be negative at higher redshifts (primarily due to $H_0$ measurement by Riess et al and the Lyman-$\alpha$ measurement of BAO at $z \sim 2.4$) has been also confirmed by several earlier works. It is, however, useful to reanalyze the data using other parametrizations and verify that the final features and results do not depend on the parametrization used in any crucial way.\\

\item[(II)] \textbf{Consistency with Planck's measurement of the CMB anisotropy:} 
\vskip 2mm

\begin{figure}
\begin{center} 
\resizebox{220pt}{220pt}{\includegraphics{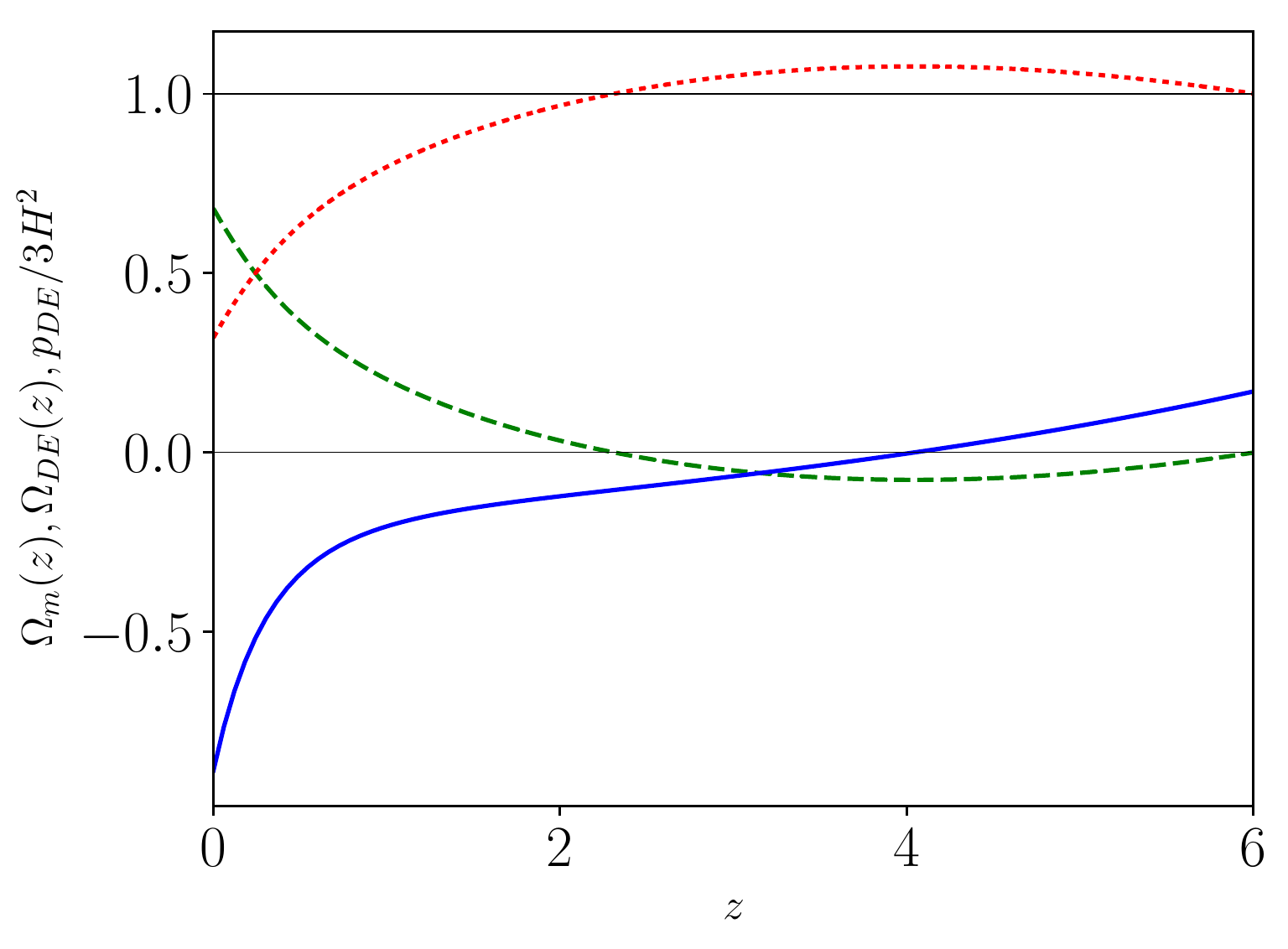}} 
\end{center}
\caption{The dotted red line is for $\Omega_{\text{m}}(z)$, the dashed green line is for $\Omega_{_{\text{DE}}}(z)$ and the solid blue line is for $\pde/(3H^2)$. This is for Planck H(z) data at $z=4,5,6$ and $\Omega^{(0)}_{m}=0.32$.  }\label{fig:omplot}
\end{figure}

The simplest dark energy model we can suggest based on our results is a negative Cosmological Constant and a phantom crossing for the rest of the dark energy. To show that such a model is consistent with CMB temperature anisotropy as measured by Planck, we calculate $C_{l}^{TT}$ for CMB anisotropy spectra assuming that till $z\leq 6$, $H(z)$ is given by best fit of our reconstructed $H(z)$ and for $z>6$, it is given by Planck best fit \lcdm model \cite{Ade:2015xua, Ade:2015rim} . Note that at $z=6$, \lcdm model is well within matter dominated era. In Fig. \ref{fig:omplot}, we show the behaviours of density parameters for matter and dark energy. As one can see, around $z=6$, $\Omega_{m} = 1$ and $\Omega_{\text{DE}} = 0$, allowing us to match our reconstructed $H(z)$ with a matter dominated era. 
\vskip 1.5mm

Using this $H(z)$, we use CLASS code \cite{class} to compute the $C_{l}^{TT}$ and compare with Planck data as well as Planck best fit \lcdm model. The result is shown in Fig.\ref{fig:CMB-ell}. As expected and one can see, $C_{l}^{TT}$ in our model is a good fit to Planck's measurement.\\

\item[(III)] \textbf{Effects on Structure Formation:} 
\vskip 2mm

The existence of this small negative  $\Lambda$ can have interesting effect on growth of structures. As one can see in Fig. \ref{fig:omplot}, $\Omega_{m}$ is slightly greater than $1$ for a certain redshift range depending on where we match the $H(z)$ with Planck's \lcdm model. This will give enhancement in growth of structures at higher redshifts and the nonlinear regime may start earlier than  in \lcdm model. This may result in the presence of more massive galaxies at higher redshifts compared to $\Lambda$CDM model, effects on reionization process as well as on lensing. All these are potential observable signatures for our model that can be tested by present and next generation galaxy surveys and CMB experiments.\\

\item[(IV)] \textbf{Modeling the $\rde (z)$:} 
\vskip 2mm

Within our data analysis framework, we have a clear indication that {dark energy sector cannot be described by {only} a cosmological constant}. Moreover, as discussed,  $\rde, \pde$ cannot be obtained from a minimally coupled scalar field (with any potential); that is, quintessence models do not lead to our $\rde(z)$. Given that $\rde$ takes negative values for a range of $z$, the simplest model is to assume presence of a \emph{negative cosmological constant} with its value $\Lambda=\rho_{\text{min}}$ and then try to model $\rho=\rde-\rho_{\text{min}}$ within a non-minimally coupled scalar theory with a positive definite potential. This latter should be such that it provides  crossing to  phantom region $(\rde+\pde <0)$ for $z<\zm$. Such models can be constructed, e.g. within Brans-Dicke theory \cite{Wang:2018fng}. Seeking and exploring such models is postponed to future works. 
\\

\item[(V)] \textbf{Theoretical implications of our dark energy model:} 
\vskip 2mm

A positive cosmological constant which is assumed to drive the current accelerated expansion of the Universe is a theoretical challenge: Getting a vacuum solution with a positive cosmological constant within moduli-fixed consistent and stable string theory compactifications has been a daunting task \cite{Maldacena:2000mw, Kachru:2003aw, Conlon:2007gk, Danielsson:2018ztv}. Moreover, formulating quantum field theory on the background of a de Sitter space has its own challenges, from the choice of the vacuum state to non-existence of a well-defined S-matrix (on global de Sitter space) \cite{Witten:2001kn, Goheer:2002vf}. Our findings here, lifts all those questions by simply removing the need for a positive cosmological constant. 
\vskip 1.5mm

\begin{figure}
\begin{center} 
\resizebox{220pt}{220pt}{\includegraphics{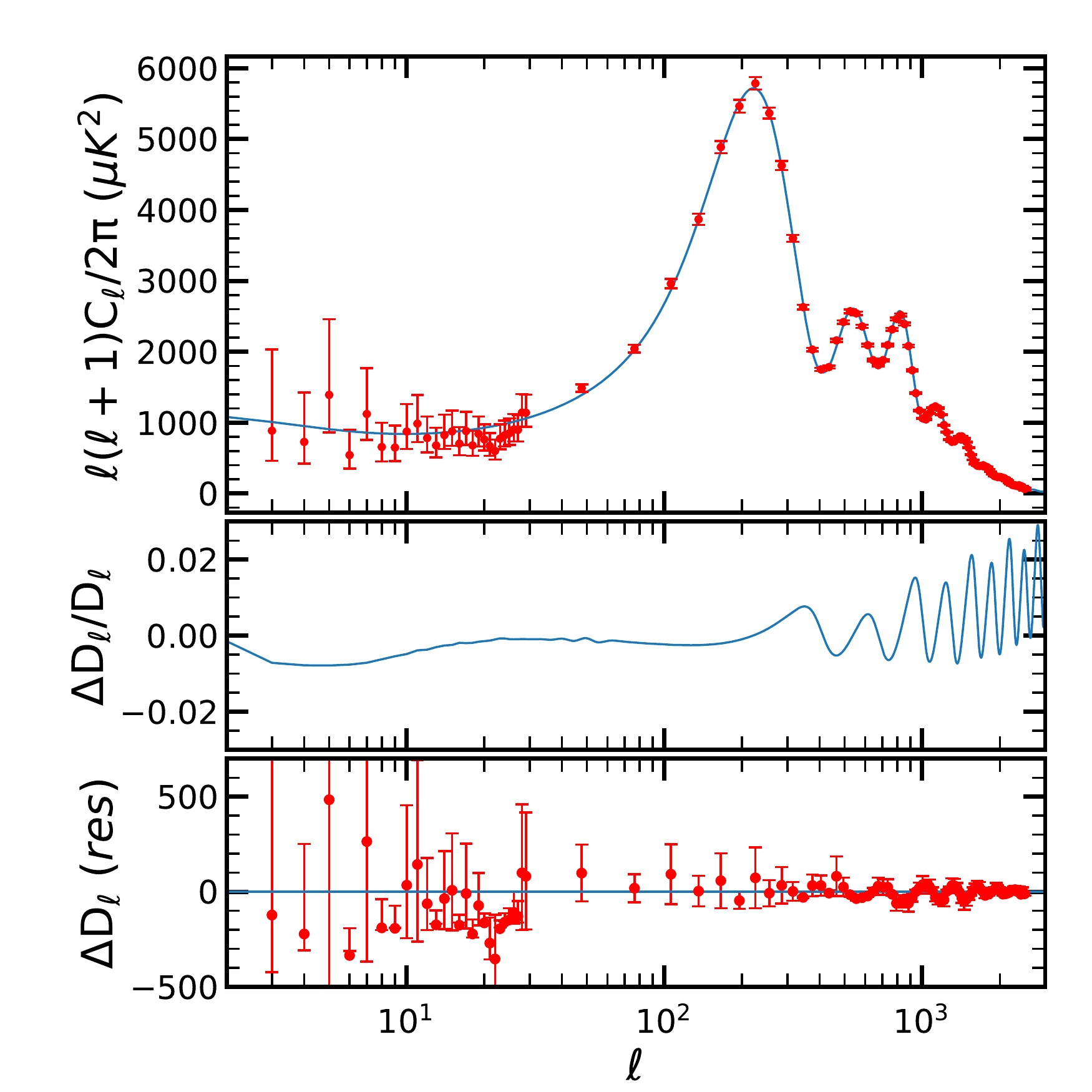}} 
\end{center}
\caption{CMB TT spectra. Top one: for model considered in this work (see text) together with Planck data and error bars for TT spectra. Middle one: the difference in TT spectra for our model and Planck best fit $\Lambda$CDM model. Bottom one: The residual for our model with the Planck data. }\label{fig:CMB-ell}
\end{figure}

On the other hand, a negative cosmological constant is a theoretical sweet spot: it provides an anti-de Sitter (AdS)  background, which is very much welcome due to AdS/CFT duality \cite{Maldacena:1997re}, providing a ``dual'' framework for  cosmology. In addition, string theory clearly prefers AdS background to de Sitter, consistent AdS backgrounds are ubiquitous in string theory settings \cite{BP}. Interestingly, it has been argued that accelerated expansion of the Universe may be possible with a negative cosmological constant \cite{HHH}. 
\vskip 1.5mm

Finally, we comment on the recently proposed ``swampland'' conjecture \cite{swampland} which favors quintessence models over $\Lambda$CDM. Our results here are in clear tension with the swampland conjecture. Similar statement has also been made in \cite{Eoin}.
\\
\item[(VI)] \textbf{Our results and anthropic reasoning.}
\vskip 2mm

In mid 1980's and long before observational establishment of late-time cosmic acceleration, based on structure formation constraints which is necessary to yield existence of life (as it is usually perceived) it was argued that the value of cosmological constant, if positive, should not be much bigger than $H_0^2$ \cite{Weinberg:1987}. A negative value of the cosmological constant, too, is bounded by similar anthropic reasoning \cite{Barrow-Tipler}.  Interestingly, the negative cosmological constant in our model which is within $1\%$ of the current total energy density of the Universe is certainly consistent with these bounds. 
\end{itemize}

{We emphasize that we ascribe the tension between the inferred value of $H_0$ between the local measurements and the Planck data fully on the dynamics of dark energy. Given this is the true case for this tension, what we find is that a negative value of the cosmological constant is still allowed by the data.} To summarise, we discuss the prospects of solving the tension between low  and high redshift cosmological observations in the presence of a small {negative cosmological constant}. This is probably the first time, that there are observational suggestions for presence of a {negative cosmological constant}. Its presence predicts definite observational signatures in large scale structure formation in the Universe and can be tested with present and future experiments.

\textbf{Note added:} As we were updating and revising our paper, the paper \cite{Riess:2019cxk} appeared which has now a new measured value for $H_{0}$ which is $74.03 \pm 1.42$ km/s/Mpc. This is now in tension with Planck-2018 measurement for $H_{0}$ at $4.4\sigma$. For our case with PL2 ($z_{match} = 6$), the value for $H_{0}$ is $70^{+1.45}_{-1.50}$ Km/s/Mpc. The deviation of this value with \cite{Riess:2019cxk} is $1.98\sigma$ which is at less than $2\sigma$ tension. {See also \cite{Dutta:2019pio} for more discussions and analysis. }

\section*{Acknowledgements}
We are grateful to Ravi Sheth and Fernando Quevedo  for fruitful discussions and Eoin  O'Colgain, Hossein Yavartanoo, Maurice van Putten for comments on the draft.
MMShJ was supported by the grants from ICTP NT-04, INSF junior chair in black hole physics, grant No 950124. AS, MMShJ  and KD would like to thank the hospitality of the Abdus-Salam ICTP, Italy, where this project  was initiated. Ruchika acknowledges the funding from CSIR, Govt. of India under Junior Research Fellowship.

\providecommand{\href}[2]{#2}\begingroup\raggedright\endgroup


\begin{thebibliography}{10}

\bibitem{Ade:2015xua}
{\bf Planck} Collaboration, P.~A.~R. Ade {\em et al.}, ``{Planck 2015 results.
  XIII. Cosmological parameters},'' {\em Astron. Astrophys.} {\bf 594} (2016)
  A13,
\href{http://www.arXiv.org/abs/1502.01589}{{\tt 1502.01589}}.
%%CITATION = ARXIV:1502.01589;%%.

\bibitem{Ade:2015rim}
{\bf Planck} Collaboration, P.~A.~R. Ade {\em et al.}, ``{Planck 2015 results.
  XIV. Dark energy and modified gravity},'' {\em Astron. Astrophys.} {\bf 594}
  (2016) A14,
\href{http://www.arXiv.org/abs/1502.01590}{{\tt 1502.01590}}.
%%CITATION = ARXIV:1502.01590;%%.

\bibitem{Riess:2016jrr}
A.~G. Riess {\em et al.}, ``{A 2.4\% Determination of the Local Value of the
  Hubble Constant},'' {\em Astrophys. J.} {\bf 826} (2016), no.~1, 56,
\href{http://www.arXiv.org/abs/1604.01424}{{\tt 1604.01424}}.
%%CITATION = ARXIV:1604.01424;%%.

\bibitem{Riess:2017lxs}
A.~G. Riess {\em et al.}, ``{Type Ia Supernova Distances at Redshift >1.5 from
  the Hubble Space Telescope Multi-cycle Treasury Programs: The Early Expansion
  Rate},'' {\em Astrophys. J.} {\bf 853} (2018), no.~2, 126,
\href{http://www.arXiv.org/abs/1710.00844}{{\tt 1710.00844}}.
%%CITATION = ARXIV:1710.00844;%%.

\bibitem{Riess:2018byc}
A.~G. Riess {\em et al.}, ``{Milky Way Cepheid Standards for Measuring Cosmic
  Distances and Application to Gaia DR2: Implications for the Hubble
  Constant},'' {\em Astrophys. J.} {\bf 861} (2018), no.~2, 126,
\href{http://www.arXiv.org/abs/1804.10655}{{\tt 1804.10655}}.
%%CITATION = ARXIV:1804.10655;%%.

\bibitem{Akrami:2018vks}
{\bf Planck} Collaboration, Y.~Akrami and others (Planck 2018~Collaboration),
  ``{Planck 2018 results. I. Overview and the cosmological legacy of Planck},''
\href{http://www.arXiv.org/abs/1807.06205}{{\tt 1807.06205}}.
%%CITATION = ARXIV:1807.06205;%%.

\bibitem{Aghanim:2018eyx}
{\bf Planck} Collaboration, N.~Aghanim and others (Planck 2018~Collaboration),
  ``{Planck 2018 results. VI. Cosmological parameters},''
\href{http://www.arXiv.org/abs/1807.06209}{{\tt 1807.06209}}.
%%CITATION = ARXIV:1807.06209;%%.

\bibitem{Bonvin:2016crt}
V.~Bonvin {\em et al.}, ``{H0LiCOW - V. New COSMOGRAIL time delays of HE
  0435-1223: H0 to 3.8 per cent precision from strong lensing in a flat LCDM
  model},'' {\em Mon. Not. Roy. Astron. Soc.} {\bf 465} (2017), no.~4,
  4914--4930,
\href{http://www.arXiv.org/abs/1607.01790}{{\tt 1607.01790}}.
%%CITATION = ARXIV:1607.01790;%%.

\bibitem{Delubac:2014aqe}
{\bf BOSS} Collaboration, T.~Delubac {\em et al.}, ``{Baryon acoustic
  oscillations in the Ly-alpha forest of BOSS DR11 quasars},'' {\em Astron.
  Astrophys.} {\bf 574} (2015) A59,
\href{http://www.arXiv.org/abs/1404.1801}{{\tt 1404.1801}}.
%%CITATION = ARXIV:1404.1801;%%.

\bibitem{Sahni:2014ooa}
V.~Sahni, A.~Shafieloo, and A.~A. Starobinsky, ``{Model independent evidence
  for dark energy evolution from Baryon Acoustic Oscillations},'' {\em
  Astrophys. J.} {\bf 793} (2014), no.~2, L40,
\href{http://www.arXiv.org/abs/1406.2209}{{\tt 1406.2209}}.
%%CITATION = ARXIV:1406.2209;%%.

\bibitem{Sola:2018sjf}
J.~Sola~Peracaula, A.~Gomez-Valent, and J.~d.~C. Perez, ``{Signs of Dynamical
  Dark Energy in Current Observations},'' {\em Phys. Dark Univ.} {\bf 25}
  (2019) 100311,
\href{http://www.arXiv.org/abs/1811.03505}{{\tt 1811.03505}}.
%%CITATION = ARXIV:1811.03505;%%.

\bibitem{Sola:2016ecz}
J.~Solà~Peracaula, J.~de~Cruz~Pérez, and A.~Gómez-Valent, ``{Dynamical dark
  energy vs. $\Lambda$ = const in light of observations},'' {\em EPL} {\bf 121}
  (2018), no.~3, 39001,
\href{http://www.arXiv.org/abs/1606.00450}{{\tt 1606.00450}}.
%%CITATION = ARXIV:1606.00450;%%.

\bibitem{Ryan:2018aif}
J.~Ryan, S.~Doshi, and B.~Ratra, ``{Constraints on dark energy dynamics and
  spatial curvature from Hubble parameter and baryon acoustic oscillation
  data},'' {\em Mon. Not. Roy. Astron. Soc.} {\bf 480} (2018), no.~1, 759--767,
\href{http://www.arXiv.org/abs/1805.06408}{{\tt 1805.06408}}.
%%CITATION = ARXIV:1805.06408;%%.

\bibitem{Ooba:2018dzf}
J.~Ooba, B.~Ratra, and N.~Sugiyama, ``{Planck 2015 constraints on
  spatially-flat dynamical dark energy models},''
\href{http://www.arXiv.org/abs/1802.05571}{{\tt 1802.05571}}.
%%CITATION = ARXIV:1802.05571;%%.

\bibitem{DiValentino:2017zyq}
E.~Di~Valentino, A.~Melchiorri, E.~V. Linder, and J.~Silk, ``{Constraining Dark
  Energy Dynamics in Extended Parameter Space},'' {\em Phys. Rev.} {\bf D96}
  (2017), no.~2, 023523,
\href{http://www.arXiv.org/abs/1704.00762}{{\tt 1704.00762}}.
%%CITATION = ARXIV:1704.00762;%%.

\bibitem{Rivera:2016zzr}
A.~Bonilla~Rivera and J.~G. Farieta, ``{Exploring the Dark Universe:
  constraints on dynamical Dark Energy models from CMB, BAO and growth rate
  measurements},'' {\em Int. J. Mod. Phys.} {\bf D28} (2019), no.~09, 1950118,
\href{http://www.arXiv.org/abs/1605.01984}{{\tt 1605.01984}}.
%%CITATION = ARXIV:1605.01984;%%.

\bibitem{Zhao:2017cud}
G.-B. Zhao {\em et al.}, ``{Dynamical dark energy in light of the latest
  observations},'' {\em Nat. Astron.} {\bf 1} (2017), no.~9, 627--632,
\href{http://www.arXiv.org/abs/1701.08165}{{\tt 1701.08165}}.
%%CITATION = ARXIV:1701.08165;%%.

\bibitem{Zhang:2017idq}
Y.~Zhang, H.~Zhang, D.~Wang, Y.~Qi, Y.~Wang, and G.-B. Zhao, ``{Probing
  dynamics of dark energy with latest observations},'' {\em Res. Astron.
  Astrophys.} {\bf 17} (2017), no.~6, 050,
\href{http://www.arXiv.org/abs/1703.08293}{{\tt 1703.08293}}.
%%CITATION = ARXIV:1703.08293;%%.

\bibitem{Poulin:2018zxs}
V.~Poulin, K.~K. Boddy, S.~Bird, and M.~Kamionkowski, ``{Implications of an
  extended dark energy cosmology with massive neutrinos for cosmological
  tensions},'' {\em Phys. Rev.} {\bf D97} (2018), no.~12, 123504,
\href{http://www.arXiv.org/abs/1803.02474}{{\tt 1803.02474}}.
%%CITATION = ARXIV:1803.02474;%%.

\bibitem{Wang:2018fng}
Y.~Wang, L.~Pogosian, G.-B. Zhao, and A.~Zucca, ``{Evolution of dark energy
  reconstructed from the latest observations},'' {\em Astrophys. J.} {\bf 869}
  (2018) L8,
\href{http://www.arXiv.org/abs/1807.03772}{{\tt 1807.03772}}.
%%CITATION = ARXIV:1807.03772;%%.

\bibitem{Shafieloo:2007cs}
A.~Shafieloo, ``{Model Independent Reconstruction of the Expansion History of
  the Universe and the Properties of Dark Energy},'' {\em Mon. Not. Roy.
  Astron. Soc.} {\bf 380} (2007) 1573--1580,
\href{http://www.arXiv.org/abs/astro-ph/0703034}{{\tt astro-ph/0703034}}.
%%CITATION = ASTRO-PH/0703034;%%.

\bibitem{Seikel:2012uu}
M.~Seikel, C.~Clarkson, and M.~Smith, ``{Reconstruction of dark energy and
  expansion dynamics using Gaussian processes},'' {\em JCAP} {\bf 1206} (2012)
  036,
\href{http://www.arXiv.org/abs/1204.2832}{{\tt 1204.2832}}.
%%CITATION = ARXIV:1204.2832;%%.

\bibitem{Gerardi:2019obr}
F.~Gerardi, M.~Martinelli, and A.~Silvestri, ``{Reconstruction of the Dark
  Energy equation of state from latest data: the impact of theoretical
  priors},'' {\em JCAP} {\bf 1907} (2019) 042,
\href{http://www.arXiv.org/abs/1902.09423}{{\tt 1902.09423}}.
%%CITATION = ARXIV:1902.09423;%%.

\bibitem{Wang:2019ufm}
D.~Wang, W.~Zhang, and X.-H. Meng, ``{Searching for the evidence of dynamical
  dark energy},'' {\em Eur. Phys. J.} {\bf C79} (2019), no.~3, 211,
\href{http://www.arXiv.org/abs/1903.08913}{{\tt 1903.08913}}.
%%CITATION = ARXIV:1903.08913;%%.

\bibitem{Saini:1999ba}
T.~D. Saini, S.~Raychaudhury, V.~Sahni, and A.~A. Starobinsky,
  ``{Reconstructing the cosmic equation of state from supernova distances},''
  {\em Phys. Rev. Lett.} {\bf 85} (2000) 1162--1165,
\href{http://www.arXiv.org/abs/astro-ph/9910231}{{\tt astro-ph/9910231}}.
%%CITATION = ASTRO-PH/9910231;%%.

\bibitem{Gruber:2013wua}
C.~Gruber and O.~Luongo, ``{Cosmographic analysis of the equation of state of
  the universe through Pade approximations},'' {\em Phys. Rev.} {\bf D89}
  (2014), no.~10, 103506,
\href{http://www.arXiv.org/abs/1309.3215}{{\tt 1309.3215}}.
%%CITATION = ARXIV:1309.3215;%%.

\bibitem{Wei:2013jya}
H.~Wei, X.-P. Yan, and Y.-N. Zhou, ``{Cosmological Applications of Pade
  Approximant},'' {\em JCAP} {\bf 1401} (2014) 045,
\href{http://www.arXiv.org/abs/1312.1117}{{\tt 1312.1117}}.
%%CITATION = ARXIV:1312.1117;%%.

\bibitem{Rezaei:2017yyj}
M.~Rezaei, M.~Malekjani, S.~Basilakos, A.~Mehrabi, and D.~F. Mota,
  ``{Constraints to Dark Energy Using PADE Parameterizations},'' {\em
  Astrophys. J.} {\bf 843} (2017), no.~1, 65,
\href{http://www.arXiv.org/abs/1706.02537}{{\tt 1706.02537}}.
%%CITATION = ARXIV:1706.02537;%%.

\bibitem{Mehrabi:2018oke}
A.~Mehrabi and S.~Basilakos, ``{Dark energy reconstruction based on the Pade
  approximation; an expansion around the $\varLambda $ CDM},'' {\em Eur. Phys.
  J.} {\bf C78} (2018), no.~11, 889,
\href{http://www.arXiv.org/abs/1804.10794}{{\tt 1804.10794}}.
%%CITATION = ARXIV:1804.10794;%%.

\bibitem{Benetti:2019gmo}
M.~Benetti and S.~Capozziello, ``{Connecting early and late epochs by f(z)CDM
  cosmography},''
\href{http://www.arXiv.org/abs/1910.09975}{{\tt 1910.09975}}.
%%CITATION = ARXIV:1910.09975;%%.

\bibitem{Capozziello:2018jya}
S.~Capozziello, Ruchika, and A.~A. Sen, ``{Model independent constraints on
  dark energy evolution from low-redshift observations},'' {\em Mon. Not. Roy.
  Astron. Soc.} {\bf 484} (2019) 4484,
\href{http://www.arXiv.org/abs/1806.03943}{{\tt 1806.03943}}.
%%CITATION = ARXIV:1806.03943;%%.

\bibitem{Evslin:2017qdn}
J.~Evslin, A.~A. Sen, and Ruchika, ``{Price of shifting the Hubble constant},''
  {\em Phys. Rev.} {\bf D97} (2018), no.~10, 103511,
\href{http://www.arXiv.org/abs/1711.01051}{{\tt 1711.01051}}.
%%CITATION = ARXIV:1711.01051;%%.

\bibitem{Gomez-Valent:2018hwc}
A.~Gomez-Valent and L.~Amendola, ``{H0 from cosmic chronometers and Type Ia
  supernovae, with Gaussian Processes and the novel Weighted Polynomial
  Regression method},'' {\em JCAP} {\bf 1804} (2018), no.~04, 051,
\href{http://www.arXiv.org/abs/1802.01505}{{\tt 1802.01505}}.
%%CITATION = ARXIV:1802.01505;%%.

\bibitem{Pinho:2018unz}
A.~M. Pinho, S.~Casas, and L.~Amendola, ``{Model-independent reconstruction of
  the linear anisotropic stress $\eta$},'' {\em JCAP} {\bf 1811} (2018),
  no.~11, 027,
\href{http://www.arXiv.org/abs/1805.00027}{{\tt 1805.00027}}.
%%CITATION = ARXIV:1805.00027;%%.

\bibitem{Reid:2012hm}
M.~J. Reid, J.~A. Braatz, J.~J. Condon, K.~Y. Lo, C.~Y. Kuo, C.~M.~V.
  Impellizzeri, and C.~Henkel, ``{The Megamaser Cosmology Project: IV. A Direct
  Measurement of the Hubble Constant from UGC 3789},'' {\em Astrophys. J.} {\bf
  767} (2013) 154,
\href{http://www.arXiv.org/abs/1207.7292}{{\tt 1207.7292}}.
%%CITATION = ARXIV:1207.7292;%%.

\bibitem{Kuo:2012hg}
C.~Kuo, J.~A. Braatz, M.~J. Reid, F.~K.~Y. Lo, J.~J. Condon, C.~M.~V.
  Impellizzeri, and C.~Henkel, ``{The Megamaser Cosmology Project. V. An
  Angular Diameter Distance to NGC 6264 at 140 Mpc},'' {\em Astrophys. J.} {\bf
  767} (2013) 155,
\href{http://www.arXiv.org/abs/1207.7273}{{\tt 1207.7273}}.
%%CITATION = ARXIV:1207.7273;%%.

\bibitem{Gao:2015tqd}
F.~Gao, J.~A. Braatz, M.~J. Reid, K.~Y. Lo, J.~J. Condon, C.~Henkel, C.~Y. Kuo,
  C.~M.~V. Impellizzeri, D.~W. Pesce, and W.~Zhao, ``{The Megamaser Cosmology
  Project VIII. A Geometric Distance to NGC 5765b},'' {\em Astrophys. J.} {\bf
  817} (2016), no.~2, 128,
\href{http://www.arXiv.org/abs/1511.08311}{{\tt 1511.08311}}.
%%CITATION = ARXIV:1511.08311;%%.

\bibitem{DiValentino:2019qzk}
E.~Di~Valentino, A.~Melchiorri, and J.~Silk, ``{Planck evidence for a closed
  Universe and a possible crisis for cosmology},'' {\em Nat. Astron.} (2019)
\href{http://www.arXiv.org/abs/1911.02087}{{\tt 1911.02087}}.
%%CITATION = ARXIV:1911.02087;%%.

\bibitem{Carvalho:2015ica}
G.~C. Carvalho, A.~Bernui, M.~Benetti, J.~C. Carvalho, and J.~S. Alcaniz,
  ``{Baryon Acoustic Oscillations from the SDSS DR10 galaxies angular
  correlation function},'' {\em Phys. Rev.} {\bf D93} (2016), no.~2, 023530,
\href{http://www.arXiv.org/abs/1507.08972}{{\tt 1507.08972}}.
%%CITATION = ARXIV:1507.08972;%%.

\bibitem{Alcaniz:2016ryy}
J.~S. Alcaniz, G.~C. Carvalho, A.~Bernui, J.~C. Carvalho, and M.~Benetti,
  ``{Measuring baryon acoustic oscillations with angular two-point correlation
  function},'' {\em Fundam. Theor. Phys.} {\bf 187} (2017) 11--19,
\href{http://www.arXiv.org/abs/1611.08458}{{\tt 1611.08458}}.
%%CITATION = ARXIV:1611.08458;%%.

\bibitem{Carvalho:2017tuu}
G.~C. Carvalho, A.~Bernui, M.~Benetti, J.~C. Carvalho, E.~de~Carvalho, and
  J.~S. Alcaniz, ``{Measuring the transverse baryonic acoustic scale from the
  SDSS DR11 galaxies},''
\href{http://www.arXiv.org/abs/1709.00271}{{\tt 1709.00271}}.
%%CITATION = ARXIV:1709.00271;%%.

\bibitem{plchain}
ESA Cosmological Data webpage on wikipedia,
  \href{https://wiki.cosmos.esa.int/planckpla/index.php/Cosmological_Parameters}{\tt{https://wiki.cosmos.esa.int/}}.

\bibitem{Csaki:2004ha}
C.~Csaki, N.~Kaloper, and J.~Terning, ``{Exorcising w <-1 },'' {\em Annals
  Phys.} {\bf 317} (2005) 410--422,
\href{http://www.arXiv.org/abs/astro-ph/0409596}{{\tt astro-ph/0409596}}.
%%CITATION = ASTRO-PH/0409596;%%.

\bibitem{Csaki:2005vq}
C.~Csaki, N.~Kaloper, and J.~Terning, ``{The Accelerated acceleration of the
  Universe},'' {\em JCAP} {\bf 0606} (2006) 022,
\href{http://www.arXiv.org/abs/astro-ph/0507148}{{\tt astro-ph/0507148}}.
%%CITATION = ASTRO-PH/0507148;%%.

\bibitem{Riess:2019cxk}
A.~G. Riess, S.~Casertano, W.~Yuan, L.~M. Macri, and D.~Scolnic, ``{Large
  Magellanic Cloud Cepheid Standards Provide a 1\% Foundation for the
  Determination of the Hubble Constant and Stronger Evidence for Physics beyond
  $\Lambda$CDM},'' {\em Astrophys. J.} {\bf 876} (2019), no.~1, 85,
\href{http://www.arXiv.org/abs/1903.07603}{{\tt 1903.07603}}.
%%CITATION = ARXIV:1903.07603;%%.

\bibitem{Aviles:2014rma}
A.~Aviles, A.~Bravetti, S.~Capozziello, and O.~Luongo, ``{Precision cosmology
  with Pade rational approximations: Theoretical predictions versus
  observational limits},'' {\em Phys. Rev.} {\bf D90} (2014), no.~4, 043531,
\href{http://www.arXiv.org/abs/1405.6935}{{\tt 1405.6935}}.
%%CITATION = ARXIV:1405.6935;%%.

\bibitem{class}
D.~Blas, J.~Lesgourgues, and T.~Tram, ``{The Cosmic Linear Anisotropy Solving
  System (CLASS) II: Approximation schemes},'' {\em JCAP} {\bf 1107} (2011)
  034,
\href{http://www.arXiv.org/abs/1104.2933}{{\tt 1104.2933}}.
%%CITATION = ARXIV:1104.2933;%%.

\bibitem{Maldacena:2000mw}
J.~M. Maldacena and C.~Nunez, ``{Supergravity description of field theories on
  curved manifolds and a no go theorem},'' {\em Int. J. Mod. Phys.} {\bf A16}
  (2001) 822--855, \href{http://www.arXiv.org/abs/hep-th/0007018}{{\tt
  hep-th/0007018}}.
[,182(2000)].
%%CITATION = HEP-TH/0007018;%%.

\bibitem{Kachru:2003aw}
S.~Kachru, R.~Kallosh, A.~D. Linde, and S.~P. Trivedi, ``{De Sitter vacua in
  string theory},'' {\em Phys. Rev.} {\bf D68} (2003) 046005,
\href{http://www.arXiv.org/abs/hep-th/0301240}{{\tt hep-th/0301240}}.
%%CITATION = HEP-TH/0301240;%%.

\bibitem{Conlon:2007gk}
J.~P. Conlon and F.~Quevedo, ``{Astrophysical and cosmological implications of
  large volume string compactifications},'' {\em JCAP} {\bf 0708} (2007) 019,
\href{http://www.arXiv.org/abs/0705.3460}{{\tt 0705.3460}}.
%%CITATION = ARXIV:0705.3460;%%.

\bibitem{Danielsson:2018ztv}
U.~H. Danielsson and T.~Van~Riet, ``{What if string theory has no de Sitter
  vacua?},'' {\em Int. J. Mod. Phys.} {\bf D27} (2018), no.~12, 1830007,
\href{http://www.arXiv.org/abs/1804.01120}{{\tt 1804.01120}}.
%%CITATION = ARXIV:1804.01120;%%.

\bibitem{Witten:2001kn}
E.~Witten, ``{Quantum gravity in de Sitter space},'' in {\em {Strings 2001:
  International Conference Mumbai, India, January 5-10, 2001}}.
\newblock 2001.
\newblock
\href{http://www.arXiv.org/abs/hep-th/0106109}{{\tt hep-th/0106109}}.
\newblock
%%CITATION = HEP-TH/0106109;%%.

\bibitem{Goheer:2002vf}
N.~Goheer, M.~Kleban, and L.~Susskind, ``{The Trouble with de Sitter space},''
  {\em JHEP} {\bf 07} (2003) 056,
\href{http://www.arXiv.org/abs/hep-th/0212209}{{\tt hep-th/0212209}}.
%%CITATION = HEP-TH/0212209;%%.

\bibitem{Maldacena:1997re}
J.~M. Maldacena, ``{The Large N limit of superconformal field theories and
  supergravity},'' {\em Int. J. Theor. Phys.} {\bf 38} (1999) 1113--1133,
  \href{http://www.arXiv.org/abs/hep-th/9711200}{{\tt hep-th/9711200}}.
[Adv. Theor. Math. Phys.2,231(1998)].
%%CITATION = HEP-TH/9711200;%%.

\bibitem{BP}
R.~Bousso and J.~Polchinski, ``{Quantization of four form fluxes and dynamical
  neutralization of the cosmological constant},'' {\em JHEP} {\bf 06} (2000)
  006,
\href{http://www.arXiv.org/abs/hep-th/0004134}{{\tt hep-th/0004134}}.
%%CITATION = HEP-TH/0004134;%%.

\bibitem{HHH}
J.~B. Hartle, S.~W. Hawking, and T.~Hertog, ``{Accelerated Expansion from
  Negative $\Lambda$},''
\href{http://www.arXiv.org/abs/1205.3807}{{\tt 1205.3807}}.
%%CITATION = ARXIV:1205.3807;%%.

\bibitem{swampland}
P.~Agrawal, G.~Obied, P.~J. Steinhardt, and C.~Vafa, ``{On the Cosmological
  Implications of the String Swampland},'' {\em Phys. Lett.} {\bf B784} (2018)
  271--276,
\href{http://www.arXiv.org/abs/1806.09718}{{\tt 1806.09718}}.
%%CITATION = ARXIV:1806.09718;%%.

\bibitem{Eoin}
E.~O. Colgain, M.~H. P.~M. Van~Putten, and H.~Yavartanoo, ``{Observational
  consequences of $H_0$ tension in de Sitter Swampland},''
\href{http://www.arXiv.org/abs/1807.07451}{{\tt 1807.07451}}.
%%CITATION = ARXIV:1807.07451;%%.

\bibitem{Weinberg:1987}
S.~Weinberg, ``{Anthropic Bound on the Cosmological Constant},'' {\em Phys.
  Rev. Lett.} {\bf 59} (1987)
2607.
%%CITATION = PRLTA,59,2607;%%.

\bibitem{Barrow-Tipler}
J.~D. Barrow and F.~J. Tipler, {\em {The Anthropic Cosmological Principle}}.
\newblock Oxford U. Pr., Oxford,
1988.
\newblock
%%CITATION = INSPIRE-1267732;%%.

\bibitem{Dutta:2019pio}
K.~Dutta, A.~Roy, Ruchika, A.~A. Sen, and M.~M. Sheikh-Jabbari, ``{Cosmology
  with low-redshift observations: No signal for new physics},'' {\em Phys.
  Rev.} {\bf D100} (2019), no.~10, 103501,
\href{http://www.arXiv.org/abs/1908.07267}{{\tt 1908.07267}}.
%%CITATION = ARXIV:1908.07267;%%.

\end{thebibliography}
\end{document}